\documentclass[a4paper,11pt]{article}

\usepackage[utf8]{inputenc}   
\usepackage{newunicodechar}   
\DeclareUnicodeCharacter{2217}{\ast}
\DeclareUnicodeCharacter{2212}{-}

\usepackage[english]{babel}

\usepackage[a4paper,top=2cm,bottom=2cm,left=3cm,right=3cm,marginparwidth=1.5cm]{geometry}

\usepackage{amsmath}    
\usepackage{graphicx}
\usepackage[colorlinks=true, allcolors=green, linktocpage]{hyperref}  
\usepackage{csquotes}     

\usepackage[sorting=none]{biblatex}
\usepackage{authblk}
\usepackage{cuted}
\usepackage[version=4]{mhchem}
\usepackage{xparse}
\usepackage{xr-hyper}

\begin{document}

\title{AI-Driven Phase Identification from X-ray Hyperspectral Imaging of cycled Na-ion Cathode Materials}

\author[1,2,4]{Fay\c{c}al Adrar}
\author[1,2,4]{Nicolas Folastre}
\author[1,2,6]{Chlo\'e Pablos}
\author[3]{\\Stefan Stanescu}
\author[3]{Sufal Swaraj} 
\author[1,2]{Raghvender Raghvender} 
\author[1]{Fran\c{c}ois Cadiou} 
\author[2,4,6]{Laurence Croguennec} 
\author[5,*]{Matthieu Bugnet}
\author[1,2,4,*]{Arnaud Demorti\`ere}

\affil[1]{Laboratoire de R\'eactivit\'e et de Chimie des Solides (LRCS), CNRS UMR 7314, Universit\'e de Picardie Jules Verne, Hub de l’Energie, 15 Rue Baudelocque, Amiens, France}
\affil[2]{R\'eseau sur le Stockage Electrochimique de l’Energie (RS2E), CNRS FR 3459, Hub de l’Energie, 15 Rue Baudelocque, Amiens, France}
\affil[3]{Synchrotron SOLEIL, L’Orme des Merisiers, 91190 Saint-Aubin, France}
\affil[4]{ALISTORE-European Research Institute, CNRS FR 3104, Hub de l’Energie, Rue Baudelocque, Amiens, France}
\affil[5]{CNRS, INSA Lyon, Universit\'e Claude Bernard Lyon 1, MATEIS, UMR 5510, 69621 Villeurbanne, France}
\affil[6]{Univ. Bordeaux, CNRS, Bordeaux INP, ICMCB, UMR 5026, F-33600 Pessac, France}

\affil[*]{corresponding author: matthieu.bugnet@cnrs.fr, arnaud.demortiere@cnrs.fr}

\renewcommand\Affilfont{\itshape\small}

\maketitle

\section{Abstract}
Na-ion batteries have emerged as viable candidates for large-scale energy storage applications due to resource abundance and cost advantages. The constraints imposed on their performance and durability, for instance, by complex phase transformations in positive electrode materials during electrochemical cycling, can be addressed and are thus not detrimental to their development. However, diffusion-limited Na-ion transport can drive spatially heterogeneous phase nucleation and propagation, leading to multiphase coexistence and locally non-uniform electrochemical activity, generating complex reaction pathways that challenge both mechanistic understanding and predictive material optimization. These challenges can be addressed by investigating single-crystalline regions of materials, i.e. down to the scale of individual particles, although such analyses are often constrained by energetically and/or spatially sparse hyperspectral datasets. Here, we developed an AI-driven method to process hyperspectral data under sparse sampling conditions and generate multiphase maps with nanometer-scale resolution over a micrometer-scale field of view. We applied this processing on scanning transmission X-ray microscopy (STXM) data to determine the distribution and coexistence of phases in individual particles of $\textrm{Na}_x\textrm{V}_2(\textrm{PO}_4)_2\textrm{F}_3$ cathode materials, at different states of charge, i.e. $x$ = 3.0, 2.4, 2.0, 1.0, and 1-$y \; (0.1 < y < 0.5)$. The methodology relies on a workflow which combines a Gaussian mixture variational autoencoder (GMVAE) algorithm with the Pearson correlation coefficient to identify the sodium content and map their spatial distribution. Our approach reveals nanoscale phase heterogeneity and evolution within individual particles, and improves the reliability of phase detection by identifying ambiguity zones, false assignments, and transition phases localized at grain boundaries. This study demonstrates the relevance and efficiency of coupling advanced VAE-based statistical modeling with modern hyperspectral imaging techniques to interpret phase transformations in energy materials prone to structural changes.

\textbf{Keywords:} deep learning, Pearson correlation coefficient, Gaussian mixture variational autoencoder, Na-ion battery cathode, $\textrm{Na}_3\textrm{V}_2(\textrm{PO}_4)_2\textrm{F}_3$, STXM, hyperspectral data

\newpage
\section{Introduction}

The transition to a sustainable energy landscape requires integrating renewable energy sources with efficient and scalable storage technologies \cite{hassan2024,mathiesen2015}. Lithium-ion batteries currently dominate energy storage due to their high energy density and long cycle life \cite{tarascon2001,nitta2014,vikstrom2013}. However, their large-scale use in stationary storage is limited by the high cost, limited availability, and environmental impact of critical raw metals such as lithium, nickel, and cobalt \cite{jaradat2025,shannak2024}. These challenges have stimulated growing interest in sodium-ion (Na-ion) batteries as a promising alternative, benefiting from the elemental abundance and low cost of sodium while offering comparable electrochemical performance \cite{deshmukh2025}. Among Na-ion positive electrode materials (called cathodes in the following), Na$_3$V$_2$(PO$_4$)$_2$F$_3$ (NVPF) has attracted particular attention due to its high cycling stability, elevated operating voltage, and excellent rate capability compared to sodium-based layered oxides and other polyanionic compounds \cite{yan2019,wang2025}. Despite these advantages, NVPF cathodes are not immune to degradations, which can arise from structural and interfacial instabilities \cite{essehli2020,sun2024,pianta2021,huang2022}.

A key challenge in Na-ion battery development lies in the strong interplay between electrochemical processes, mechanical effects and diffusion mechanisms, manifested through complex phase transformations within electrode materials \cite{yabuuchi2014}, which are further influenced by microstructural factors such as grain boundaries, surface reactions, porosity and carbon percolation network’s. In Na-ion cathodes, the comparatively large ionic radius of Na$^{+}$ and its stronger electrostatic interactions with host frameworks, relative to Li$^{+}$, give rise to distinct mechanims, even in similar host structures \cite{li2024}. Layered sodium oxides frequently undergo multiple phase transitions, which can hinder fast Na$^{+}$ diffusion. In addition, their pronounced sensitivity to air exposure poses significant challenges for storage and handling. Polyanionic compounds, on the other hand, are predominantly constrained by their low intrinsic electronic conductivity, which requires them to be synthesised as carbon-coated nanoparticles ($<$100 nm in diameter for the individual particles). For instance, in Na$_3$V$_2$(PO$_4$)$_2$F$_3$, the overall volume change upon cycling is relatively small ($\sim 3\%$) \cite{yan2019}, reflecting the structural robustness of this polyanionic framework. During (de)sodiation, progressive Na$^+$ extraction drives vanadium oxidation and structural phase transitions, associated to Na$^{+}$ and charge ordering,with the bulk material largely following the equilibrium phase diagram reported in Ref. \cite{broux2018}. However, spatial heterogeneities in Na$^+$ distribution within individual particles can promote local phase coexistence, generating regions with distinct electrochemical responses that influence reaction kinetics, reversibility, and long-term nanoscale stability. To fully understand these mechanisms, it is crucial to dynamically monitor charge and discharge processes by correlating electrochemical behavior with nanoscale chemical information derived from vanadium oxidation states, which serve as sensitive indicators of structural phase evolution. While mesoscale structural studies of Na$_x$V$_2$(PO$_4$)$_2$F$_3$ have established its equilibrium phase evolution during cycling \cite{bianchini2015}, additional complexity is expected to arise at the nanoscale due to local compositional and kinetic heterogeneities. Accordingly, advanced spectroscopic analyses were employed to map phase distributions via spatial variations in vanadium and oxygen valence states, enabling direct insight into local sodium-ion diffusion mechanisms across different states of charge.

For this purpose, we employed synchrotron scanning transmission X-ray microscopy (STXM) to study Na$_3$V$_2$(PO$_4$)$_2$F$_3$ (referred to as Na$_{3.0}$VPF) and its desodiated phases. STXM is particularly well-suited for this local investigation due to its ability to provide chemical and electronic state information at the nanoscale through hyperspectral imaging, where each scan point contains a full X-ray absorption spectrum. This technique enables direct visualization of phase distributions and chemical heterogeneities within individual cathode particles. Changes in lattice parameters leading to phase transformations are closely linked to variations in X-ray absorption near edge spectra (XANES) spectra, reflecting alterations in the local vanadium environment. However, the complex phase transition behavior during electrochemical cycling poses significant challenges for conventional characterization techniques, which often lack the local sensitivity to resolve these transformations. Many studies report the use of synchrotron STXM to investigate battery materials, focusing on phase transformations and electrocatalytic reaction performance during cycling \cite{mefford2021,lim2016,ohmer2015,yoo2020,askari2020,desmit2008}. Notably, Ohmer \textit{et al.}. \cite{ohmer2015} tracked phase boundary propagation in single-crystalline LiFePO$_4$ during lithiation and delithiation using \textit{in situ} STXM. A comprehensive overview of such studies can be found in the review of Kim \textit{et al.} \cite{kim2023}, and the relevance of STXM for battery research in the review of Temprano \textit{et al.} \cite{temprano2024}. Despite these advances, high-resolution mapping remains limited due to the intrinsic difficulty to disentangle the subtle and overlapping spectral features associated to each phase, thus restricting the ability to fully resolve nanoscale heterogeneities within individual particles.

STXM produces large and complex datasets that require specialized analysis tools. Traditional softwares such as aXis2000 \cite{hitchcock2023}, STXM Reader \cite{marcus2023}, and MANTiS \cite{lerotic2014} are powerful for general data processing. However, common phase-mapping approaches such as least-squares linear combination fitting (LS-LCF) become less reliable when spectral resolution is low \cite{umemoto2023}, and singular value decomposition (SVD), often used for phase mapping \cite{santos2022}, can also fail under sparse spectral sampling. This limitation often arises in high spatial resolution STXM experiments, where fewer energy points are collected to minimize radiation damage and acquisition time, resulting in sparse hyperspectral datasets. Such datasets, characterized by limited spectral sampling, challenge conventional unmixing methods because subtle spectral variations cannot be robustly captured or distinguished \cite{zhu2023,zang2024}. To overcome these limitations, we developed a custom Python-based approach combining Pearson correlation coefficient \cite{pearson1896} (PCC) and Gaussian mixture variational autoencoder (GMVAE)  methods \cite{dilokthanakul2017,kingma2013,wetzel2017,banko2021}.
Variational autoencoder models are increasingly used to analyze high-dimensional scientific imaging and hyperspectral datasets, including electron microscopy \cite{biswas2023}, liquid-phase TEM (LPTEM) \cite{shabeeb2025}, and diffraction mapping \cite{calvat2025}. Physics-augmented VAE frameworks have been shown to disentangle latent physical factors and delineate microstructural boundaries in hyperspectral microscopy \cite{biswas2023}, transformer-VAE architectures capture stochastic nanoparticle dynamics in LPTEM using physics-informed loss functions \cite{shabeeb2025}. Similar latent-space encodings have been used to derive reduced diffraction representations that reveal microstructure heterogeneity and support materials design \cite{calvat2025}. These studies collectively highlight the utility of VAE-based latent representations for extracting interpretable physical structure from complex experimental data. The Gaussian mixture component in a GMVAE corrects the unimodal latent space assumption of standard VAEs by introducing a multimodal prior, enabling the model to capture clustered, heterogeneous data distributions and thus improving structured representation and class separability in the latent space \cite{dilokthanakul2017}. The Pearson correlation identifies similarities in spectral shape, while the VAE compresses data into a compact latent space where distinct phases can be more effectively separated. 

In this work, we implement a two-step PCC-GMVAE workflow tailored for sparsely sampled STXM datasets. High resolution reference spectra for five sodium contents were used to select thirteen characteristic energies for high spatial resolution mapping. Pixel spectra were first assigned using Pearson correlation, and a reliability metric based on the gap between the two highest correlation coefficients was introduced to identify ambiguous regions. To resolve these ambiguities, spectra were projected into a three dimensional GMVAE latent space and reassigned using Mahalanobis distances to phase clusters. Applied to Na$_x$V$_2$(PO$_4$)$_2$F$_3$ cathodes at multiple states of charge, this approach reveals strong intra- and inter-particle heterogeneity and demonstrates reliable multiphase mapping, establishing a robust AI-driven framework for sparse hyperspectral phase identification.

\section{Results}

\subsection{Experimental procedure}
\begin{figure}[!ht]
    \centering
    \includegraphics[width=\textwidth]{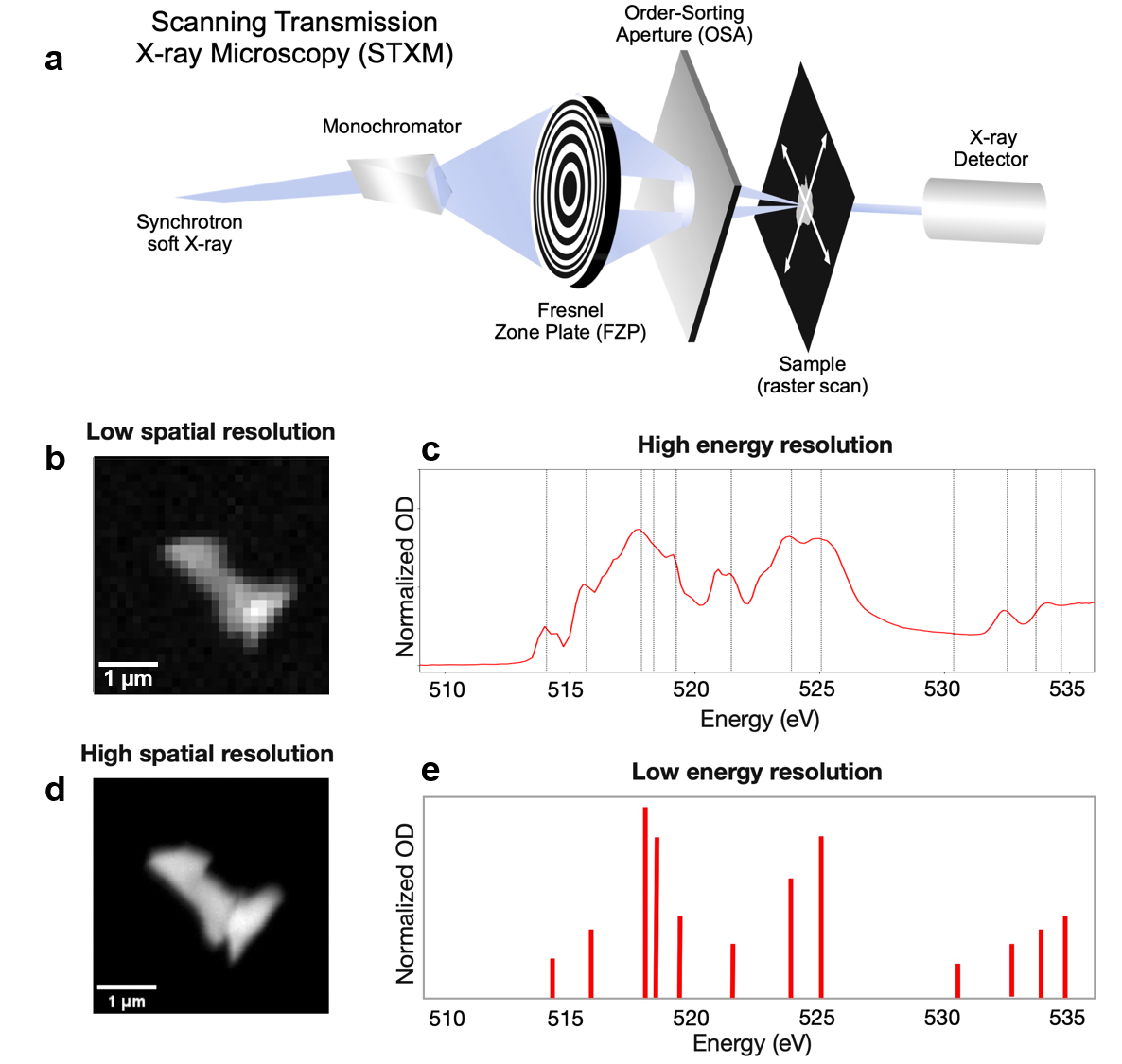}
    \caption{\small \textbf{Scanning transmission X-ray microscopy (STXM) experimental setup.} 
    (a) Schematic of the STXM setup (not to scale). 
    (b) Illustration of a stack of images acquired with low spatial resolution (low-sampling version of the image shown in panel d). 
    (c) Optical density (OD) spectrum of a NVPF particle acquired with high energy resolution; the dashed line indicates the energies selected for high spatial resolution acquisitions. 
    (d) High spatial resolution image of a NVPF particle (same as in (b)). 
    (e) Illustration of the optical density (OD) spectrum from a low spectral resolution acquisition, where the bars represent the energy levels and their heights correspond to the optical density values.}
    \label{fig:1}
\end{figure}

The experimental setup is shown in Figure \ref{fig:1}a, and a comparison between acquisitions at high spectral resolution (Figures \ref{fig:1}b--c) and high spatial resolution (Figures \ref{fig:1}d-e) highlights the intrinsic trade-off of STXM technique. High spectral resolution ($\sim$ 0.1 eV) enables the identification of detailed energy-dependent features but provides limited spatial information ($\sim$ 135 nm). Conversely, high spatial resolution ($\sim$ 31 nm) captures fine spatial details but with a reduced number of energy points, as indicated by the discrete red bars in Figures \ref{fig:1}d-e. This trade-off adds to acquisition time constraints and the risk of beam damage, making it difficult to simultaneously achieve both high spatial and spectral resolution.

To address this limitation, we first measured reference spectra at high spectral resolution for the five sodium contents. The resulting V-L$_{2,3}$ and O-K edge spectra (Figure \ref{fig:2}) exhibit strong variations at specific energies. From these data, thirteen characteristic energies were selected (black dashed lines in Figure \ref{fig:2}), which were subsequently used for high spatial resolution mapping of the Na$_x$VPF samples.

\subsection{X-ray absorption near edge spectra (XANES)}

\begin{figure}[!ht]
    \centering
    \includegraphics[width=\textwidth]{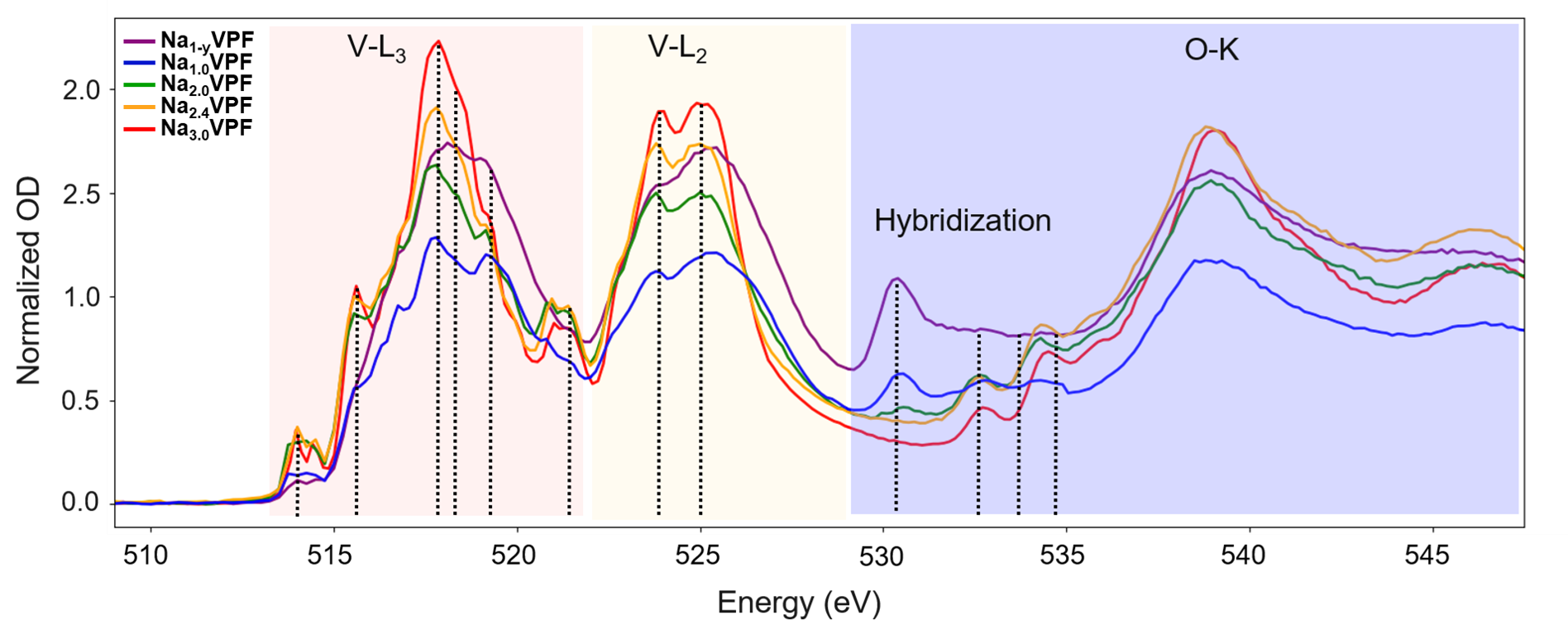}
    \caption{\small \textbf{Reference XANES spectra of NVPF for different levels of Na$^{+}$ extraction.} 
    Optical density (OD) of V-L$_{2,3}$ and O-K edge X-ray absorption spectra acquired in STXM measurements on individual NVPF particles at different states of charge. 
   }
    \label{fig:2}
\end{figure}

Figure \ref{fig:2} shows X-ray absorption near edge structure (XANES) for NVPF samples at different states of charging. The corresponding electrochemical cycling curves and X-ray diffraction patterns are presented in Supplementary Figure S1. The full pattern matching (Le Bail method) refinements results for Na$_3$VPF, Na$_{2.4}$VPF, Na$_{2}$VPF, Na$_1$VPF, and Na$_{1-y}$VPF are also provided in Supplementary Figure S2. Each spectrum corresponds to the integration over an entire particle for a given state of charge. The XANES is in good agreement with the literature \cite{yan2019} and is thus considered representative of the different Na contents.
These representative spectra from individual particles illustrate the progressive evolution of the redox state during cycling. In the pristine material, Na$_3$VPF, two intense peaks are observed at about 517.5 eV and 525.0 eV, which originate from vanadium L$_3$ (2p$_{3/2} \rightarrow$ 3d) and L$_2$ (2p$_{1/2} \rightarrow$ 3d) transitions. The splitting between these peaks arises from spin--orbit coupling of the 2p electrons \cite{yan2019,abbate1993}. As the material is deintercalated, the spectral weight of these peaks gradually shifts to higher energies, evidencing the oxidation of vanadium. This interpretation is further supported by the pre-edge features of the oxygen K-edge (530--535 eV), which correspond to transitions from O 1s to V 3d--O 2p hybridized states. The pre-edge peak at $\sim$530.3 eV becomes more pronounced during charging, indicating that the V--O bonds acquire increased covalency as vanadium loses electrons and the bonding evolves \cite{yan2019,suntivich2014}.  

The vertical dashed black lines mark twelve specific energies where significant changes in the optical density (i.e. XANES intensity) occur. An additional energy, around 505 eV, was taken before the V-L$_{2,3}$ edges, and is not visible in Figure \ref{fig:2} due to display preferences. Nevertheless, by analyzing only these thirteen energies, it is possible to extract qualitative information about the material, including the sodium content. Indeed, variations of the optical density at these energies are strongly linked to the changes in sodium levels \cite{yan2019}. In this study, these energies were used to map the phase distribution of several individual NVPF particles. Reference spectra composed of 13 selected energy points were generated for each Na content by extracting the corresponding optical density values from the high spectral resolution acquisitions shown in Figure \ref{fig:2}.

Intrinsically, this experimental approach records the spectral intensity only at selected energies and does not rely on the full XANES spectrum, as acquiring a complete high-resolution spectral stack in STXM is highly time-consuming and results in a large cumulative X-ray dose that can induce significant beam damage in sensitive battery materials. Therefore, it is primordial to develop new data processing strategies capable of capturing the subtle spectral variations and the corresponding spatial variations. The typical methods commonly used to interpret X-ray absorption fine structures, using electronic structure calculations or peak fitting, are unsuitable in this case due to the sparse energy sampling, which makes extracting quantitative information challenging \cite{fister2007,wang2021,wang2023}. Instead, approaches inspired by correlation techniques primarily enabling qualitative insights into the sample as used in diffraction pattern matching in transmission electron microscopy (e.g., ASTAR \cite{rauch2013}, PyXem \cite{cautaerts2022}, py4DSTEM \cite{ophus2021}, ePattern \cite{folastre2024}), may provide robust alternatives for analyzing such datasets.

\subsection{STXM phase mapping workflow}

\begin{figure}[!ht]
    \centering
    \includegraphics[width=\textwidth]{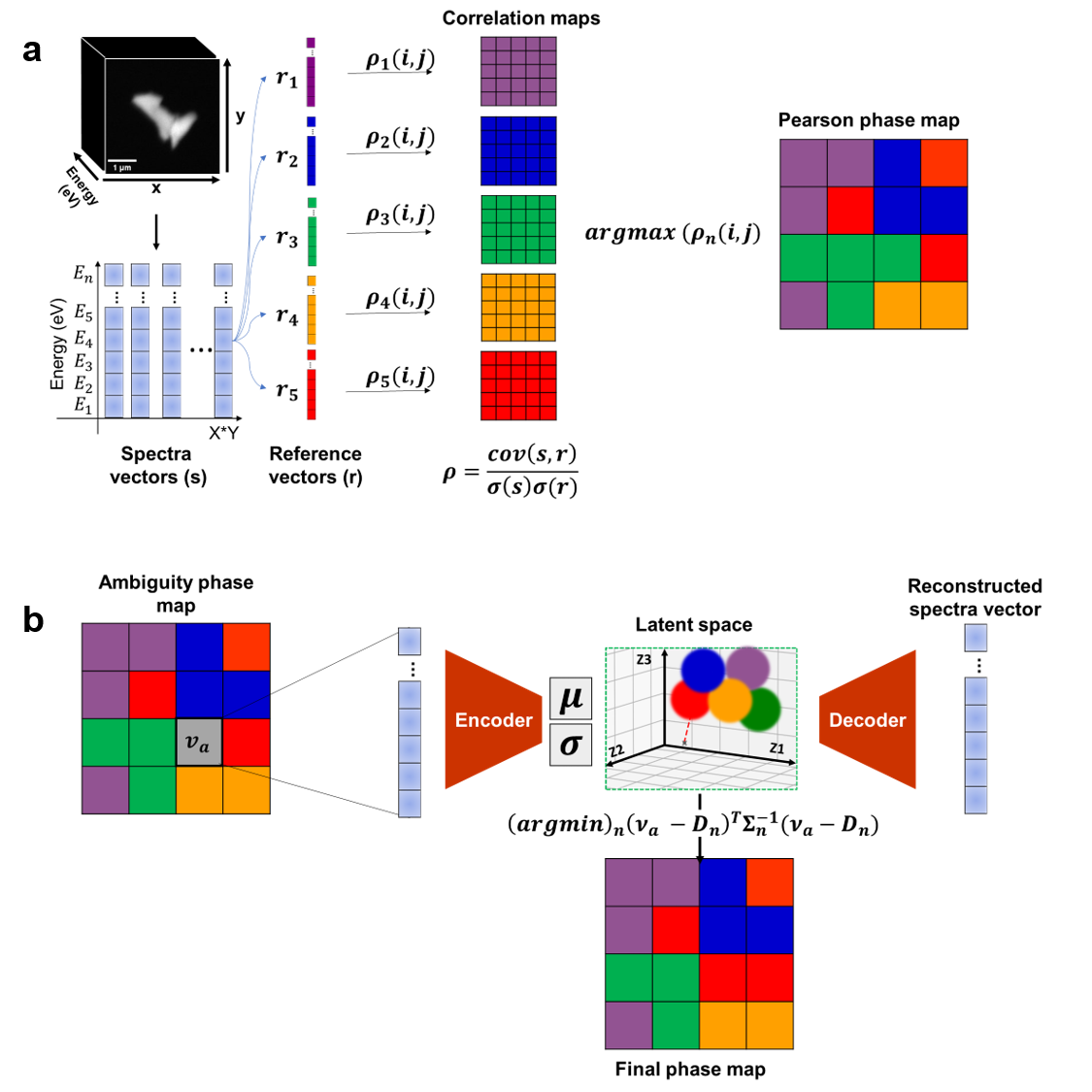}
    \caption{\small \textbf{STXM phase mapping workflow.} 
    (a) Pearson correlation phase mapping: the Pearson correlation is computed between each spectrum vector $s$ and each reference vector $r_n$ ($n$=1,2,3,4,5). 
    The resulting correlation map, $\rho_n$ ($n$=1,2,3,4,5), represents the correlation between $s$ and $r_n$ across all pixels. 
    The initial phase map is obtained by comparing the correlation values pixel by pixel and assigning each pixel to the reference with the highest correlation. 
    (b) Ambiguity map: pixels where $\arg\max(\rho_n(i,j) - \rho_m(i,j)) \leq 0.005$, with $n \ne m$, are shown in grey and labeled as ambiguous pixels $v_a$. 
    The ambiguity is resolved by projecting $v_a$ into the latent space of the trained Gaussian mixture variational autoencoder (GMVAE) containing the overall phase distributions, 
    measuring the Mahalanobis distances, and assigning the pixel to the phase corresponding to the nearest phase distribution.}
    \label{fig:3}
\end{figure}

The initial data processing was carried out using the aXis2000 software \cite{hitchcock2023}, which was employed for image alignment and normalization of the images into optical density (OD) according to the Beer--Lambert law. The extracted OD from reference samples at the characteristic energy levels provides experimental fingerprints of the sodium content in the samples.

Subsequently, we developed a Python-based workflow for phase mapping of the STXM data, as illustrated in Figure \ref{fig:3}. In this workflow, each spectrum from the spectral image data cube is compared with the five reference spectra using the PCC. This metric, highly sensitive to subtle variations in the data, quantifies the linear relationship between the measured OD spectra and the references, resulting in a correlation map for each compared sodium phase (shown in Supplementary Figure S3c--g). 

Each spectrum is then assigned to the phase with the highest PCC, yielding an initial phase map. We introduce a reliability metric $R$ tailored to this context, based on the comparison of correlation coefficients extracted from the correlation maps associated with each phase. The metric is defined as:

\begin{equation}
R = 100\left(Q_{1} - Q_{2}\right)
\end{equation}

where $R$ is the reliability score, $Q_{(1)}$ is the highest correlation coefficient among all phases for a given pixel, and $Q_{(2)}$ is the second-highest coefficient. This score quantifies how unambiguous a pixel’s phase assignment is, larger values of $R$ indicate a greater gap between the best and second-best correlations, and therefore higher confidence in the assignment. The reliability values are displayed as a greyscale map, with each pixel encoding its corresponding $R$ (Supplementary Figure S3a), which is then overlaid on the phase map as a brightness layer to simultaneously visualize both the assigned phase and the confidence of the assignment (Supplementary Figure 3b).

However, if the correlation coefficients of two phases for a given pixel differ by less than 0.005, this threshold being determined as the minimum difference required to distinguish between two reference spectra using correlation (see supplementary Figure S4), the pixel is labeled as an ambiguous phase. This designation indicates that the correlation values are too close to allow an unambiguous assignment.

To resolve spectral ambiguities, we developed a deep learning model based on a GMVAE, trained on one-dimensional XANES spectra extracted from STXM data cubes of NVPF particles at various states of charge. By incorporating a mixture of Gaussian priors, the GMVAE captures multimodal latent distributions, enabling unsupervised emergence of discrete spectral clusters corresponding to distinct chemical populations. As reported in \cite{dilokthanakul2017}, this approach improves clustering by preserving discrete structure, reducing mode mixing, and, through minimum-information constraints, producing compact and interpretable latent representations with competitive performance.

The multimodal latent prior facilitates structured organization of the latent space, allowing spectrally similar phases to be separated more robustly than with conventional correlation-based approaches. Regularization through the Kullback–Leibler divergence term in the loss function further shapes the latent distribution, grouping similar spectra into well defined clusters. This provides a more effective representation space in which ambiguous or strongly overlapping XANES features can be distinguished with higher confidence. The GMVAE architecture is shown in Supplementary Figure S5.

We designed and evaluated several strategies for phase assignment within the latent space. In the first approach, which proved to be the most robust for our dataset, all spectra are projected into the latent space and initially labeled using PCC. For each latent space cluster corresponding to a given phase, we compute the Mahalanobis distance (see Methods), which explicitly accounts for the covariance structure and anisotropy of the cluster distribution. Unlike the Euclidean distance, the Mahalanobis metric normalizes distances by the intrinsic variance of each cluster and incorporates correlations between latent variables, making it particularly well-suited for probabilistic latent spaces where clusters exhibit non-spherical geometries. Ambiguous spectra are then assigned to the phase whose latent distribution yields the minimum Mahalanobis distance, ensuring statistically consistent classification relative to the learned cluster structure. In a second approach, the experimental reference spectra are directly projected into the latent space and used as anchor points. Ambiguous spectra are then assigned to the nearest reference using Euclidean distance. This strategy is appropriate when the measured spectra have intensity scales comparable to the references. However, this condition is not strictly satisfied in our dataset, limiting its reliability. Additionally, we explored alternative assignment schemes, including K-Nearest Neighbors (KNN) and distance-to-centroid classification based on the center of mass of each phase cluster. These approaches offer flexibility in cases where cluster boundaries are less clearly separated.

\subsection{STXM phase mapping}
\begin{figure}[!ht]
    \centering
    \includegraphics[width=\textwidth]{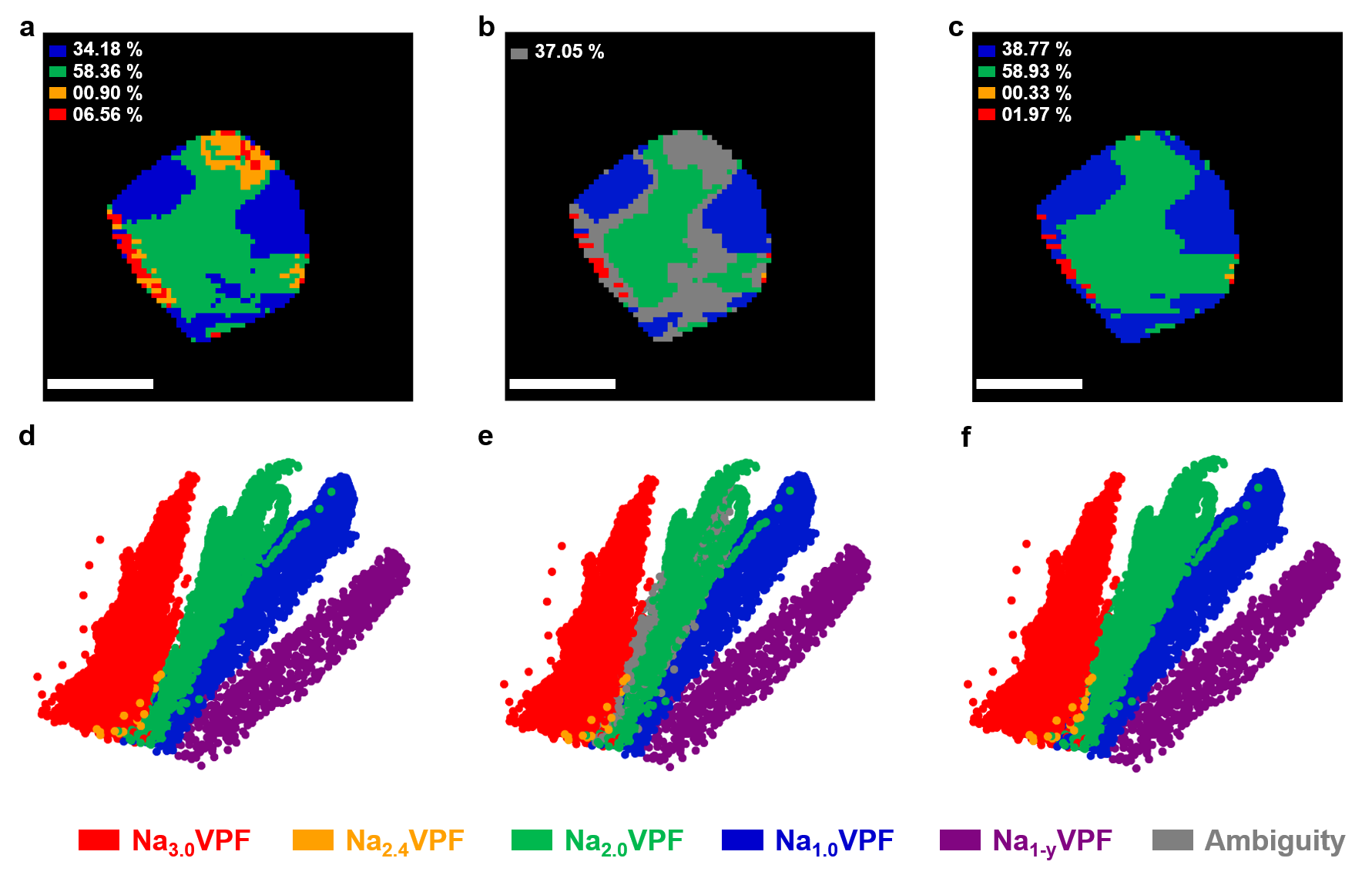} 
    \caption{\small \textbf{Phase mapping of a Na$_2$V$_2$(PO$_4$)$_2$F sample.} 
    (a) Phase map obtained using PCC
    (b) Map of ambiguous regions, where the correlation coefficient between two phases is close ($\leq 0.005$), shown in grey. 
    (c) Final phase map with ambiguities resolved using a GMVAE by projecting ambiguous spectra into the latent space and assigning them to the closest phase distribution. The scale bars correspond to $0.2\,\mu$m
    (d) Phase distributions in the GMVAE latent space corresponding to the global latent representation without the projection of ambiguous pixels. (e) Projection of ambiguous pixels (in gray). (f) Latent space after resolution of ambiguous pixels. }
    \label{fig:4}
\end{figure}

In this section, we present the phase mapping results for a Na$_2$VPF particle in the charged state. Figure \ref{fig:4}a shows the initial phase map obtained from the Pearson correlation analysis, which reveals the coexistence of four phases. Specifically, 58.36\% of the particle corresponds to the Na$_{2}$VPF phase, 34.18\% of the particle is assigned to the Na$_1$VPF phase. The Na$_3$VPF and Na$_{2.4}$VPF phase are also detected, suggesting  incomplete electrochemical transformations within the particle. Ambiguous pixels, shown in grey in Figure \ref{fig:4}b, represent about 37.05\% of the particle, reflecting uncertainty in phase assignment due to the high similarity of correlation values among the three phases.

Figure \ref{fig:4}c presents the refined phase map after resolving these ambiguities. Here, the ambiguous pixels are redistributed among the coexisting phases, yielding an adjusted composition of 58.93\% Na$_{2}$VPF, 38.77\% Na$_1$VPF, and 1.97\% Na$_3$VPF. The corresponding correlation maps for each sodium phase are displayed in Supplementary Figure S3c--g, with PCCs ranging from 0.05 to 1.0. These matrices quantify the agreement between experimental XANES spectra and the reference spectra of each identified phase, and are visualized as color-coded maps where stronger correlations appear as more intense colors. Notably, the Na$_{1}$VPF and Na$_2$VPF phases exhibit strong spectral similarity as shown in Figure \ref{fig:2}, which explains the phase ambiguity observed in Figure \ref{fig:4}b.

To address this ambiguity, the spectral vectors of the ambiguous pixels, together with the unambiguous spectra of the two phases in question, were projected into a latent space learned by a GMVAE. The analysis of the latent distributions (Figures \ref{fig:4}d--f) shows a clear separation of points into distinct clusters, underlining both the discriminative strength of the PCC and the ability of the GMVAE to resolve ambiguous cases. 

 When inspecting the latent space containing all particles (Supplementary Figure S6) it becomes clear that PCC based phase identification can produce false positives. For example, the PCC assigns some spectra to the Na$_3$VPF phase even though they originate from a particle in the Na$_1$VPF state of charge. By contrast, the GMVAE latent space places these spectra correctly near the Na$_1$VPF cluster, and inspection of the corresponding raw spectra confirms that they resemble Na$_1$VPF rather than pristine Na$_3$VPF. This demonstrates that the global latent space representation not only resolves ambiguous assignments but also provides a robust diagnostic for evaluating the reliability of PCC based methods.

\begin{figure}[!ht]
    \centering
    \includegraphics[width=\textwidth]{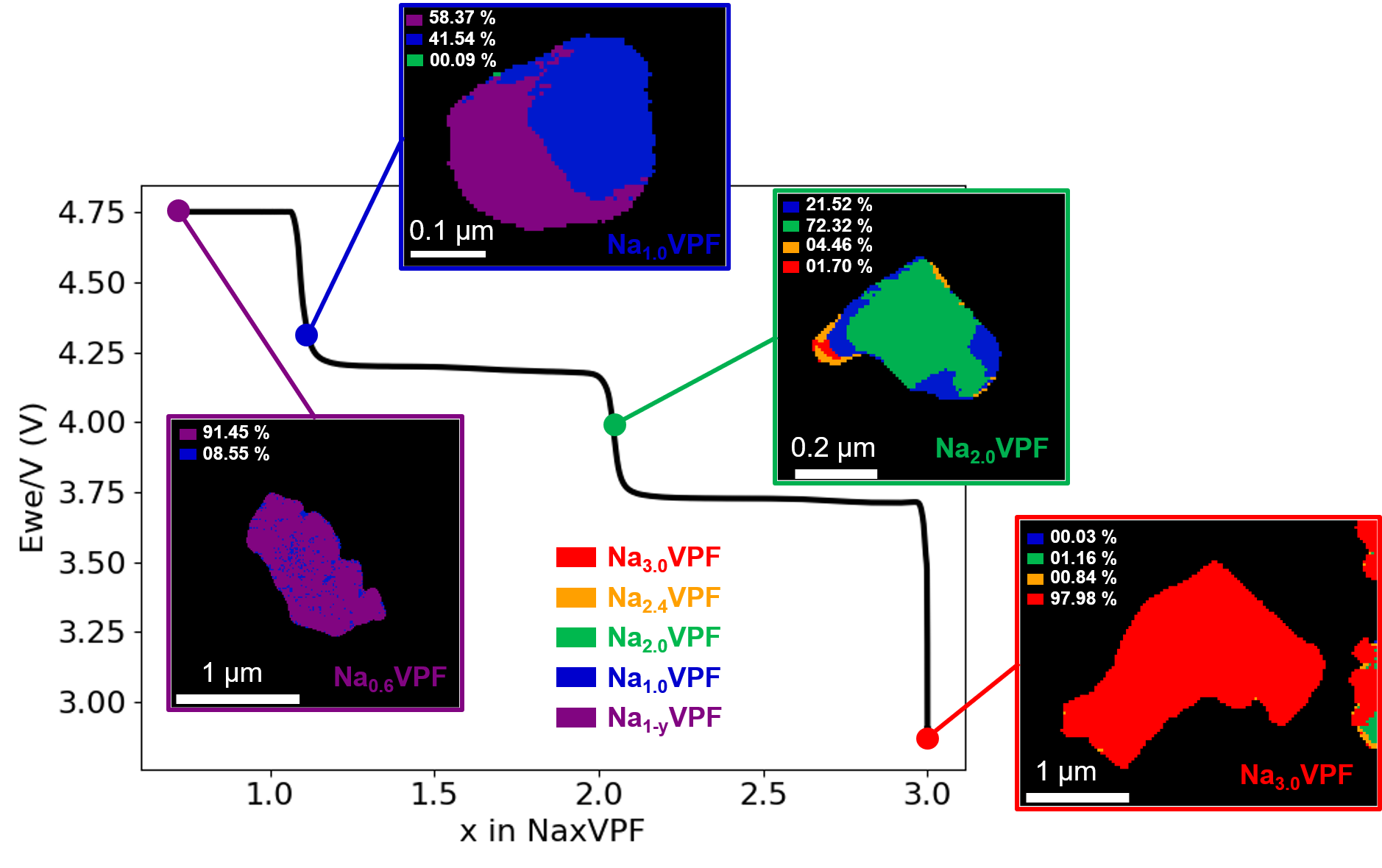} 
        \caption{\small \textbf{Phase maps of four NVPF samples at different states of charge.} 
        Phase maps of Na$_3$VPF, Na$_2$VPF, Na$_1$VPF, and Na$_{1-y}$VPF, respectively, after ambiguity resolution. Each image corresponds to a sample halted at the state of charge indicated on the electrochemical curve of NVPF.}
    \label{fig:5}
\end{figure}

Figure \ref{fig:5} presents the phase mapping results of four NVPF particles at different states of charge, ranging from Na$_{3.0}$VPF  to Na$_{1-y}$VPF. The corresponding reliability map is displayed in supplementary Figure S7. In the pristine state, the Na$_{3.0}$VPF phase dominates, consistent with the absence of desodiation. Nonetheless, traces of the Na$_{2.4}$VPF and Na$_{2}$VPF phase are detected, particularly at the particle periphery and in small fragments adjacent to the main particle. These regions exhibit low reliability in the associated reliability maps. 

At $x = 2.0$, the phase distribution becomes significantly heterogeneous, with 4.46\% Na$_{2.4}$VPF, 72.32\% Na$_{2.0}$VPF, 21.52\% Na$_{1.0}$VPF, and 1.70\% Na$_{3}$VPF.

At $x = 1.0$, three phases are identified, with 58.37\% Na$_{1-y}$VPF, 41.54\% Na$_{1.0}$VPF, and 0.09\% Na$_{2.0}$VPF. In the fully discharged state ($x = 1-y$), only two phases are detected, 91.45\% Na$_{1-y}$VPF and 8.55\% Na$_{1.0}$VPF. However, these values should be interpreted with caution. As shown in Supplementary Figure S8, the PCCs obtained for the Na$_{1.0}$VPF and Na$_{1-y}$VPF reference spectra are relatively low, reducing the confidence in the corresponding phase assignments. This limitation implies that, although the algorithm returns a dominant Na$_{1-y}$VPF contribution, the phases may not be fully reliable.

\begin{figure}[!ht]
    \centering
    \includegraphics[width=\textwidth]{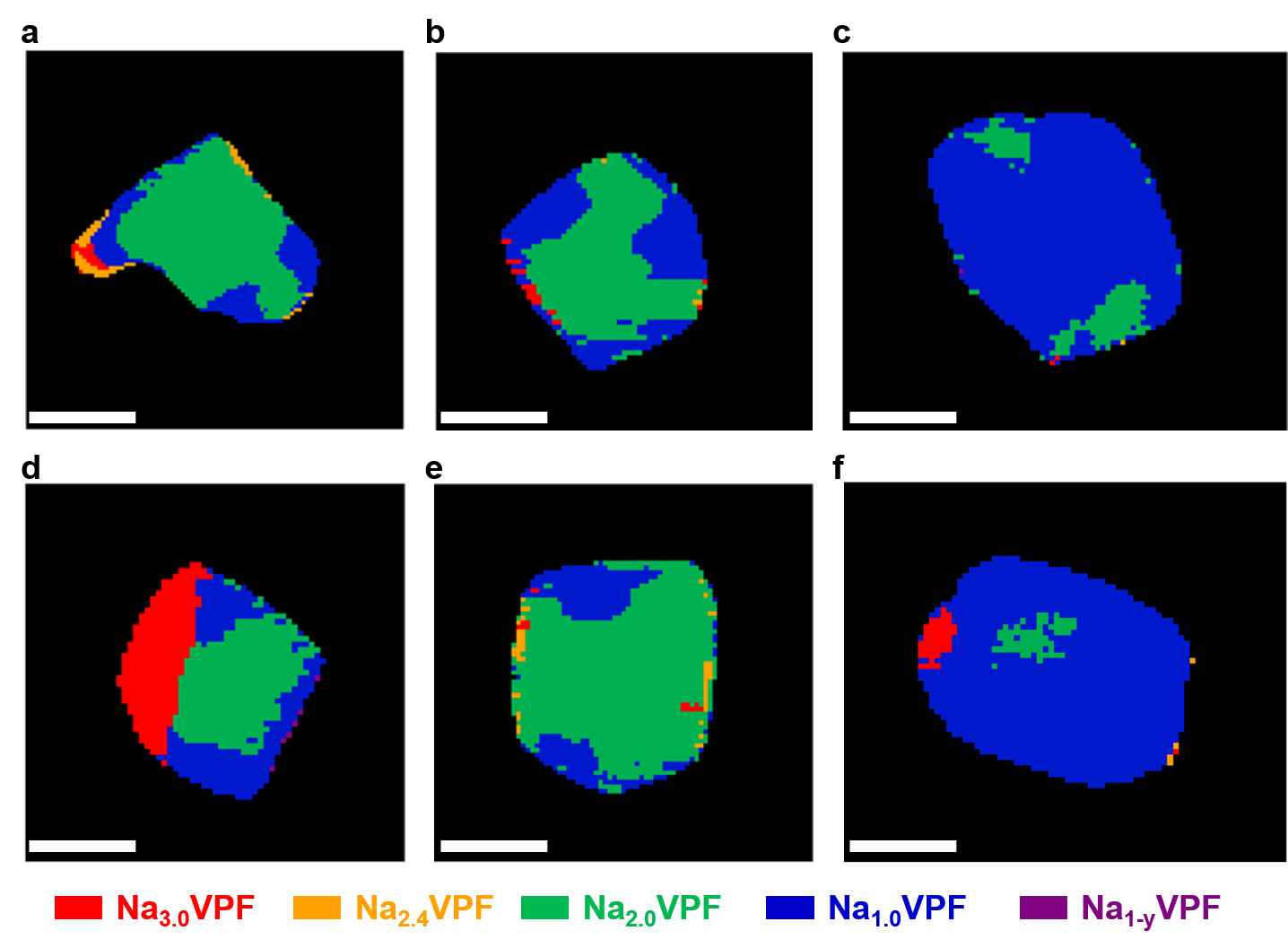}
    \caption{\small \textbf{Phase maps for six particles at the Na$_{2.0}$VPF state of charge.} The scale bar corresponds to 0.2 $\mu\text{m}$ .}
    \label{fig:6}
\end{figure}

 Our data processing framework was systematically applied to multiple particles at a single nominal state of charge to assess its robustness and reproducibility. Across the six particles examined corresponding to the state of charge Na$_{2}$VPF, we observe a highly heterogeneous and particle-dependent phase distribution, as shown in Figure \ref{fig:6}. Despite reaching this nominal state of charge, all particles contain a noticeable fraction of the Na$_{1}$VPF phase with a good reliability, indicating that they are effectively discharging faster than expected. At the same time, most particles also exhibit Na$_{2.4}$VPF and Na$_{3}$VPF phases, which shows that other regions within the same particles discharge more slowly. Overall, all six analyzed particles display a clearly heterogeneous and non-uniform phase distribution, both within individual particles and from one particle to another.

\section{Discussion}

Our phase mapping results reveal significant spatial heterogeneity in NVPF particles during desodiation. In the pristine state, the Na$_3$VPF phase predominates, confirming the absence of desodiation and validating our phase mapping workflow. However, traces of the Na$_{2.4}$VPF and Na$_{2}$VPF phases are detected at the particle periphery and in adjacent fragments, where lower reliability is observed in the associated maps. At an intermediate state of charge ($x = 2.0$), our phase maps reveal a complex coexistence of four sodium phases (Na$_{3.0}$, Na$_{2.4}$, Na$_{2.0}$, Na$_{1.0}$). The gradual evolution of Na$_3$V$_2$(PO$_4$)$_2$F$_3$ XRD patterns reported by Bianchini et al. \cite{bianchini2015} corresponds to the deintercalation process from Na$_3$VPF towards Na$_{2.4}$VPF, Na$_{2.2}$VPF and Na$_2$VPF. Although direct reference patterns for intermediate compositions such as Na$_{2.2}$VPF and Na$_{1.8}$VPF are not available in our dataset, the observed three biphasic solution may reflect a progressive and spatially asynchronous phase transformation at the nanoscale. Such heterogeneity has been attributed to partial sodium/vacancy ordering, local structural constraints, and limited ionic or electronic transport, which can lead to incomplete or delayed transitions within individual particles. In our case, we also observe that some regions of a particle discharge significantly faster than others in a strongly heterogeneous manner (Figure \ref{fig:6}c,d). While particle-to-particle variations in size and morphology could in principle influence this behavior, the particles shown in Figure \ref{fig:6} have comparable sizes, suggesting that the observed heterogeneity arises primarily from intrinsic local structural or kinetic factors rather than from geometric effects. The influence of grain orientation on the discharge heterogeneity in NVPF particles cannot be excluded, considering its significant effect on the reaction mechanisms in LiCoO$_{2}$ demonstrated in a recent study \cite{jacquet2025operando}.

Some quantitative differences are apparent when comparing our results with previous studies. While Bianchini \textit{et al.}.\cite{bianchini2015} reported that Na$_3$VPF vanishes almost completely before substantial growth of Na$_2$VPF, our phase mapping reveals persistent Na$_{3.0}$VPF and Na$_{2.4}$VPF domains. This suggests slower local transformation kinetics or stronger locally confined transport limitations. Akhtar \textit{et al.}.\cite{akhtar2023} documented structural heterogeneity and delayed transformations in NVPF cathodes, and primarily focused on material synthesis and O/F ratio control. At $x = 1.0$, only two phases are detected, with residual Na$_2$VPF, indicating that some regions remain unreacted. However, it is important to note that the reliability of this particular phase is low. Reduced reliability at phase interfaces corresponds to areas of higher strain, complicating phase discrimination \cite{santos2022}. In the fully discharged state ($x = 1-y$), only Na$_{1-y}$VPF and Na$_{1.0}$VPF remain. Overall, these high resolution phase maps demonstrate that NVPF desodiation is inherently non-uniform and multi-step, with domain-specific variations in transformation kinetics and structural constraints.

Capturing this level of detail required a Python-based phase mapping workflow, as our experimental setup involved a trade-off between achieving high spatial resolution and limiting spectral sampling to minimize beam damage. Under these conditions, conventional LS-LCF \cite{umemoto2023,lerotic2014,hitchcock2023} are unsuitable due to undersampling. Instead, we employed the PCC \cite{atoum2019,zhelezniak2019,pearson1896}, a shape-based similarity metric. To overcome cases where PCC alone cannot resolve phases with highly similar spectral signatures, we incorporated a GMVAE based ambiguity resolution  step \cite{kingma2013,wetzel2017,banko2021}. By clustering spectra in latent space, the GMVAE enables discrimination between otherwise indistinguishable phases. 

The GMVAE latent space distribution in Supplementary Figure S6a demonstrates that the model successfully captures the underlying structure of the different NVPF phases. Distinct clusters emerge, spanning from red to purple, corresponding to compositions ranging from Na$_{3}$VPF  to Na$_{1-y}$VPF. Within each cluster, the latent coordinates also reflect variations in overall spectral intensity, navigating from the upper to the lower regions of a given cluster corresponds to a systematic decrease in intensity. 

The reconstructed spectra presented in Supplementary Figure S9 further demonstrate the GMVAE’s ability to capture subtle spectroscopic evolution across the NVPF system. The model successfully reproduces the characteristic peak shifts associated with the transition from Na$_{3}$VPF to Na$_{1-y}$VPF, indicating that it has internalized key features of the V–L$_{2,3}$ edges and their evolution during sodium extraction. While these results are encouraging, they must be interpreted with caution. Beyond standard validation, it is essential to acknowledge intrinsic limitations of VAE-based models. The Kullback–Leibler regularization enforces smooth latent-space interpolation, which can inherently bias reconstructions toward averaged representations and limit the recovery of sharp or rare spectral features. As a result, VAEs may struggle to faithfully reproduce highly localized or low-probability physicochemical states, whereas generative paradigms such as diffusion models are often better suited for high-fidelity data generation. In addition, reconstruction quality remains fundamentally constrained by the size, diversity, and representativeness of the training dataset, particularly in sparsely sampled experimental regimes. Furthermore, a key next step will be to evaluate the model using synthetic spectra with controlled and physically grounded features, enabling systematic conditioning on specific experimental parameters and material states. This strategy will not only verify that the reconstructions faithfully capture intrinsic material behavior but also enhance the model’s ability to generate high-confidence outputs associated with well-defined experimental conditions and electrochemical states.

Beyond phase discrimination, the GMVAE latent space also captures physically meaningful intensity variations. As shown in Supplementary Figure S10c, the mean spectral intensity evolves systematically along the $z_2$ direction, indicating that the network organizes spectra not only by chemical phase but also by overall absorption amplitude. Since STXM optical density depends on both composition and sample thickness via the Beer–Lambert law, this trend suggests that the GMVAE implicitly clusters energy vectors as a function of effective sample thickness. This emergent structuring highlights the ability of the model to disentangle multiple experimental factors without explicit supervision, reinforcing the robustness and physical interpretability of the latent-space representation.

This combined PCC-GMVAE strategy significantly enhances the reliability and robustness of phase detection under sparse spectral sampling conditions, and highlights the capability of structured deep learning models to extract physically meaningful information from reduced hyperspectral datasets. Importantly, the GMVAE does not merely act as a clustering tool, but learns a globally consistent multimodal latent representation of the spectral landscape, enabling ambiguity resolution, false-positive correction, and statistically grounded phase discrimination. We demonstrate that even with only 13 energy acquisitions, this workflow reliably resolves multiphase distributions in charged NVPF particles, thereby overcoming the intrinsic trade-off between spatial resolution, acquisition time, and beam-induced damage in high-resolution STXM experiments.

Beyond the present case study, the GMVAE-based framework is inherently generalizable to other sparsely sampled hyperspectral modalities, including soft X-ray spectromicroscopy, EELS spectrum imaging, and STEM-EDX datasets. The latent-space formalism further opens promising perspectives for multimodal data fusion, for instance through coupling with 4D-STEM orientation and phase mapping approaches such as automated crystal orientation mapping (ACOM). Integrating spectroscopic latent encodings with diffraction-based structural descriptors would provide deeper insight into chemo-mechanical coupling and phase transformation pathways.

Future developments may involve incorporating complementary experimental techniques to improve quantitative accuracy \cite{olmos2026,su2021}, as well as extending the architecture toward deeper neural network models integrating attention mechanisms, diffusion priors, or physics-informed constraints \cite{hajimiri2021,rocha2024}. Such advances could further enhance latent space disentanglement, improve interpretability, and ultimately enable transferable AI frameworks for multimodal, beam-sensitive materials characterization.

\section{Conclusion}

In summary, this combined PCC and GMVAE strategy improves the reliability of phase detection and highlights the potential of high spatial resolution STXM for studying battery materials and other beam-sensitive systems. Notably, we show that even in the case of sparse energy sampling with only thirteen data points, the workflow can resolve desodiated phases in charged NVPF particles. More broadly, this work demonstrates that meaningful structural and chemical information can be extracted from reduced spectral datasets while maintaining nanometer-scale spatial resolution. Using NVPF at different states of charge as a case study, we demonstrate that combining high spatial resolution imaging with deep-learning-based spectral analysis provides a robust and statistically grounded framework for interpreting STXM-XANES data acquired under limited spectral sampling conditions. The proposed PCC-GMVAE workflow not only resolves phase ambiguities but also introduces a reliability metric and a latent-space-based validation strategy that strengthen confidence in phase assignment. By explicitly accounting for cluster covariance and multimodal spectral distributions, the approach enables consistent discrimination of spectrally overlapping phases and reveals pronounced intra- and inter-particle heterogeneity during desodiation. Beyond the specific NVPF system, this study establishes a generalizable AI-driven methodology for phase identification in sparsely sampled hyperspectral datasets. The framework can be extended to other energy and beam-sensitive materials and adapted to complementary modalities such as X-ray or electron diffraction. More broadly, it demonstrates how deep learning latent representations enable data-efficient and physically interpretable high-resolution chemical mapping under experimental constraints.

\section{Methods}

\subsection{STXM}
Samples of $\textrm{Na}_x\textrm{V}_2(\textrm{PO}_4)_2\textrm{F}_3$ (where $x = 3, 2.4, 2, 1, 1-y$) were analyzed using Scanning Transmission X-ray Microscopy (STXM) at the HERMES beamline of the SOLEIL synchrotron \cite{belkhou2015}. A Zone plate focusing optics, (50 nm outerzone width) was used with a phosphor coated photomultiplier tube (PMT) as a photon detector. The STXM chamber was maintained at 1$e^{-5}$ mbar pressure during the measurements. NVPF particles collected from the scratched electrode were dispersed in ultrapure ethanol and drop-cast onto an amorphous carbon-coated TEM grid, followed by solvent evaporation at room temperature. Sample preparation and handling were performed under air-protected conditions to minimize degradation induced by moisture exposure.

\subsection{Optical density}
STXM provides access to OD images at different energies, where the OD is defined as

\begin{equation}
    \text{OD} = \ln\left( \frac{I_0}{I} \right)
\end{equation}

where $I_0$ is the intensity of the incident beam, and $I$ is the transmitted beam intensity. This relationship is derived from the beer-Lambert law, expressed as

\begin{equation}
    I = I_0 \exp\left( \mu(E) \rho d \right)
\end{equation}

where $\mu(E)$ is the energy-dependent mass absorption coefficient, $\rho$ is the sample density, and $d$ is the sample thickness.

\subsection{Data pre-processing}
we started the processing of STXM data several by aligning the sequence of images, or stack, using a cross-correlation-based program called Jacobson alignment, implemented in aXis2000. This alignment step is essential to guarantee that all images are correctly registered, with the same region of interest consistently located across the entire stack. The images are then converted into OD maps using the Beer-Lambert law, which relates the transmitted X-ray intensity to the concentration of the absorbing species within the material. then to perform a phase mapping the method described in section "phase mapping workflow".
\subsection{Pearson Correlation Coefficient}
The Pearson correlation coefficient measures the linear relationship between two variables 
$X$ and $Y$, quantifying both strength and direction:
\[
r_{XY} = \frac{\mathrm{Cov}(X,Y)}{\sigma_X \, \sigma_Y}\]

where $\mathrm{Cov}(X,Y)$ is the covariance between $X$ and $Y$, and $\sigma_X$ and $\sigma_Y$ are the standard deviations of $X$ and $Y$, respectively.

The coefficient ranges from $-1$ to $1$, with $1$ indicating perfect positive linear correlation, $-1$ indicating perfect negative linear correlation, and $0$ indicating no linear relationship.

\subsection{Mahalanobis distance}
The Mahalanobis distance measures the distance between a point 
$\mathbf{x}$ and a distribution 
$\mathcal{N}(\boldsymbol{\mu}_n, \boldsymbol{\Sigma}_n)$, accounting for correlations and variable scales:

\[
D_M = (\mathbf{x} - \boldsymbol{\mu}_n)^\mathrm{T} \, \boldsymbol{\Sigma}_n^{-1} \, (\mathbf{x} - \boldsymbol{\mu}_n)
\]

It generalizes Euclidean distance by incorporating the covariance structure, providing a normalized multivariate measure.

\subsection{Gaussian mixture variational autoencoder (GMVAE)}
We implemented a GMVAE in TensorFlow/Keras to model hyperspectral STXM data and extract low-dimensional representations of the underlying phases. The dataset was randomly split into training (95\%), validation (2.5\%), and test (2.5\%) subsets to train and evaluate the model. The encoder consists of four fully connected layers with 512, 256, 128, and 64 units (ReLU activations), followed by batch normalization and dropout. It outputs the parameters of a 3-dimensional latent space together with the mixture-component probabilities defining the Gaussian mixture prior. The decoder mirrors the encoder symmetrically (64–128–256–512 units) and reconstructs the spectra through a final linear layer. The GMVAE was trained using a reconstruction loss (MSE) and KL-divergence term adapted for mixture priors.

\subsection{Material synthesis}

The synthesis of $\textrm{Na}_3\textrm{V}_2(\textrm{PO}_4)_2\textrm{F}_3$ was achieved using $\textrm{VPO}_4$ and NaF in stoechiometric amounts as precursor materials. To prepare $\textrm{VPO}_4$, stoichiometric amounts of $\textrm{V}_2\textrm{O}_5$ (Sigma-Aldrich, $\geq$99.6\%) and $\textrm{NH}_4\textrm{H}_2\textrm{PO}_4$ (Sigma-Aldrich, $\geq$99.99\%) were thoroughly mixed using a high-energy ball mill for 90 minutes. The resulting mixture was subjected to a two-step thermal treatment under a flowing Ar/H$_2$ (95\%/5\%) atmosphere to fully reduce vanadium from its +5 to +3 oxidation state: first at 300\,$^\circ$C for 5 hours with a heating rate of 0.5\,$^\circ$C/min, followed by a second step at 890\,$^\circ$C for 2 hours with a heating rate of 3\,$^\circ$C/min. After the heat treatments, the material was cooled to ambient temperature at a controlled rate of 3\,$^\circ$C/min.

\subsection{Electrochemistry}

Electrochemical cycling was employed to prepare $\textrm{Na}_x\textrm{V}_2(\textrm{PO}_4)_2\textrm{F}_3$ (NVPF) samples at different charge and discharge states. The electrodes were prepared from a mixture of NVPF (94 wt\%), carbon black (3 wt\%), and PVDF binder (3 wt\%, Sigma-Aldrich). Sodium metal was used as both the counter and reference electrode. The electrolyte consisted of 1\,M $\textrm{NaPF}_6$ dissolved in a 1:1 weight ratio of ethylene carbonate (EC) and dimethyl carbonate (DMC), with 2\,wt.\% fluoroethylene carbonate (FEC) as an additive. A Celgard membrane was used as the separator. Cells were cycled at a rate of C/10 (corresponding to the complete exchange of $\textrm{Na}^+$ per e$^-$ over 10 hours) to reach specific voltages, obtaining samples corresponding to $x = 3, 2$ (at 4.0\,V vs. $\textrm{Na}^+$), 1 (at 4.3\,V vs. $\textrm{Na}^+$), and also attempting to reach $x < 1$, stopping at $x = 1-y$ (at 4.75\,V vs. $\textrm{Na}^+$).

\subsection{Code availability}
The STXM phase mapping Python code is open source and available on GitHub at: https://github.com/Image-DataScience-Team-LRCS/stxm-phase-mapping

\subsection{Acknowledgements}
This work was supported by RS2E network (Reseau Fran\c{c}ais sur le Stockage Electrochimique de l’Energie) and the French National Research Agency under the France 2030 program (Grant ANR-22-PEBA-0002, PEPR Batteries). AD, MB, NF, FC and FA acknowledge access to beamtime and technical support from the HERMES beamline at the SOLEIL Synchrotron, France. CP and LC thank the European Union’s Horizon 2020 research and innovation program under grant agreement N 875629 (NAIMA project) for funding.
\newpage

\printbibliography

\end{document}


\maketitle
\newpage
\section{Supplementary Figures}

\FloatBarrier

Laboratory powder X-ray diffraction (PXRD) patterns were recorded on a PANalytical X'Pert 3 diffractometer in Debye--Scherrer $\theta$--$\theta$ geometry. Routine acquisitions were performed on powders packed in 0.5~mm diameter capillaries over a $2\theta$ range of $10$--$80^{\circ}$ with a step size of $0.0099^{\circ}$. The diffractometer was equipped with a Cu K$\alpha_{1,2}$ X-ray source. Full-pattern matching refinements were carried out using the Le Bail method as implemented in Jana2006~\cite{petricek2014}.

\begin{figure}[h!]
\centering
\includegraphics[width=0.75\textwidth]{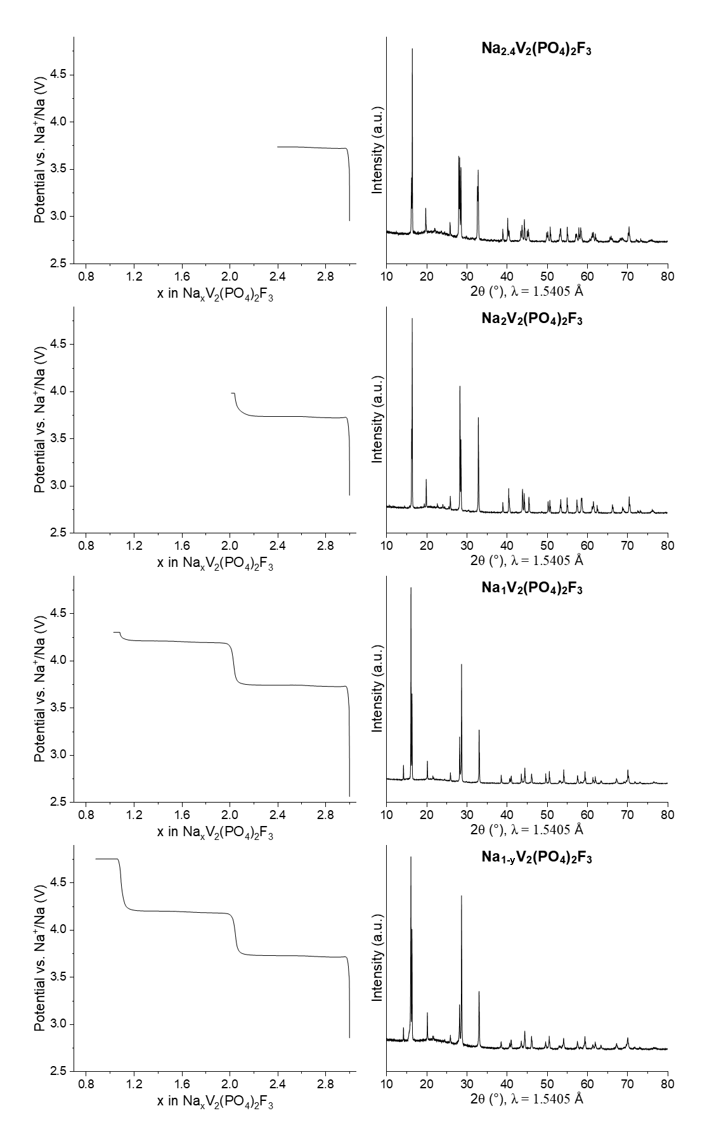}
\caption{On the right, the NVPF electrochemical curves corresponding to each state of charge investigated in this study. On the left, the associated X-ray diffraction (XRD) patterns collected at the same states of charge, highlighting the structural evolution during (de)sodiation.}
\label{fig:supp1}
\end{figure}

\FloatBarrier

\begin{figure}[h!]
\centering
\includegraphics[width=0.95\textwidth]{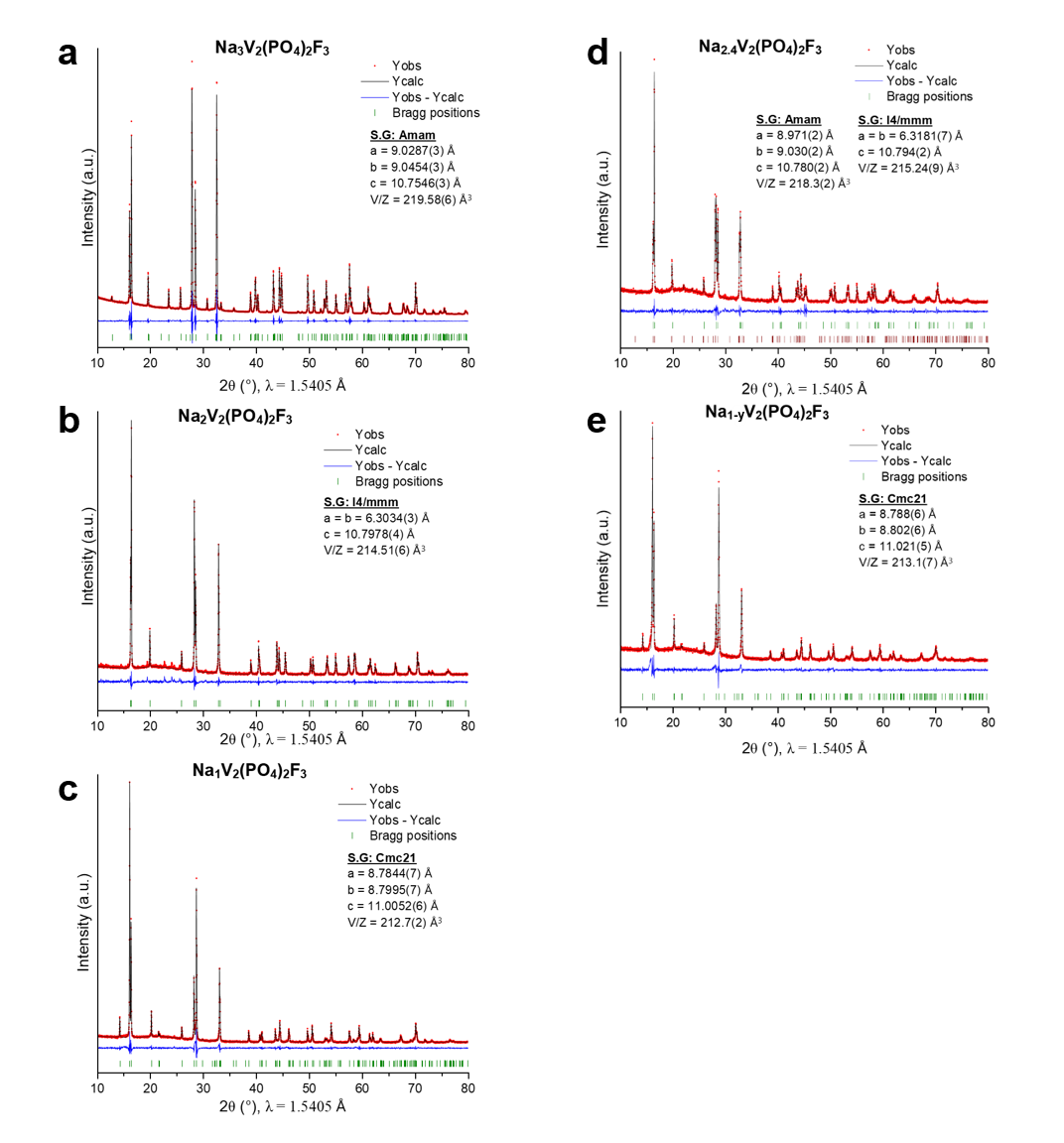}
\caption{Le Bail refinements of (a) Na$_3$V$_2$(PO$_4$)$_2$F$_3$ in the orthorhombic \textit{Amam} space group, 
(b) Na$_2$V$_2$(PO$_4$)$_2$F$_3$ in the tetragonal \textit{I4/mmm} space group,
(c) Na$_1$V$_2$(PO$_4$)$_2$F$_3$ in the orthorhombic
\textit{Cmc2$_1$} space group.
(d) Na$_2.4$V$_2$(PO$_4$)$_2$F$_3$ a mixture of phases belonging to the orthorhombic \textit{Amam} space group and the tetragonal \textit{I4/mmm} space group, and
(e) Na$_1-y$V$_2$(PO$_4$)$_2$F$_3$ in the orthorhombic
\textit{Cmc2$_1$} space group }
\label{fig:supp2}
\end{figure}

\FloatBarrier

\begin{figure}[h!]
\centering
\includegraphics[width=0.95\textwidth]{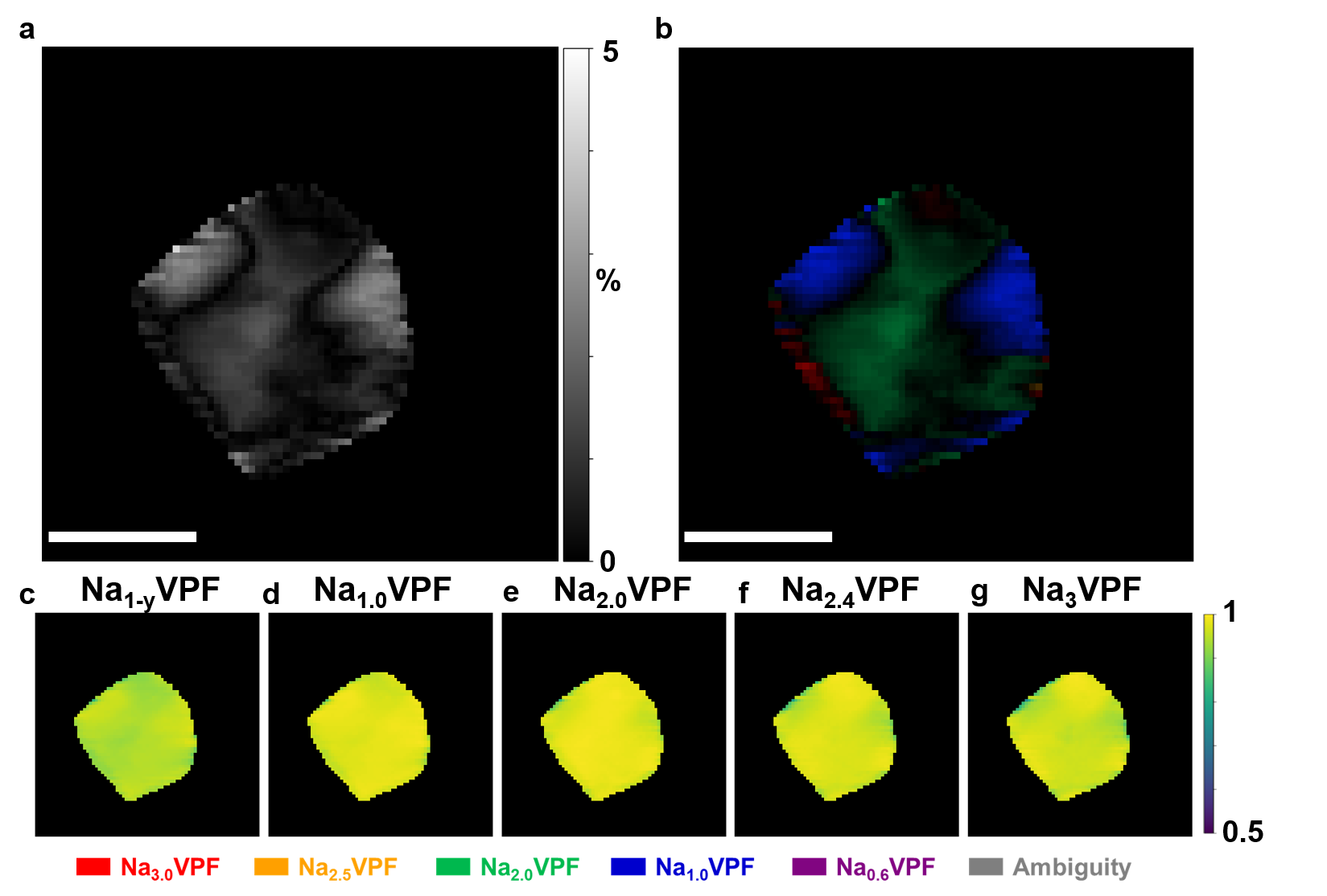}
\caption{a) Reliability maps. b) Combined reliability map with the phase map.
c--g) Correlation maps obtained for Na\textsubscript{2.0}VPF. The scale bars correspond to $0.2\,\mu m$}.
\label{fig:supp3}
\end{figure}

\begin{figure}[h!]
\centering
\includegraphics[width=0.95\textwidth]{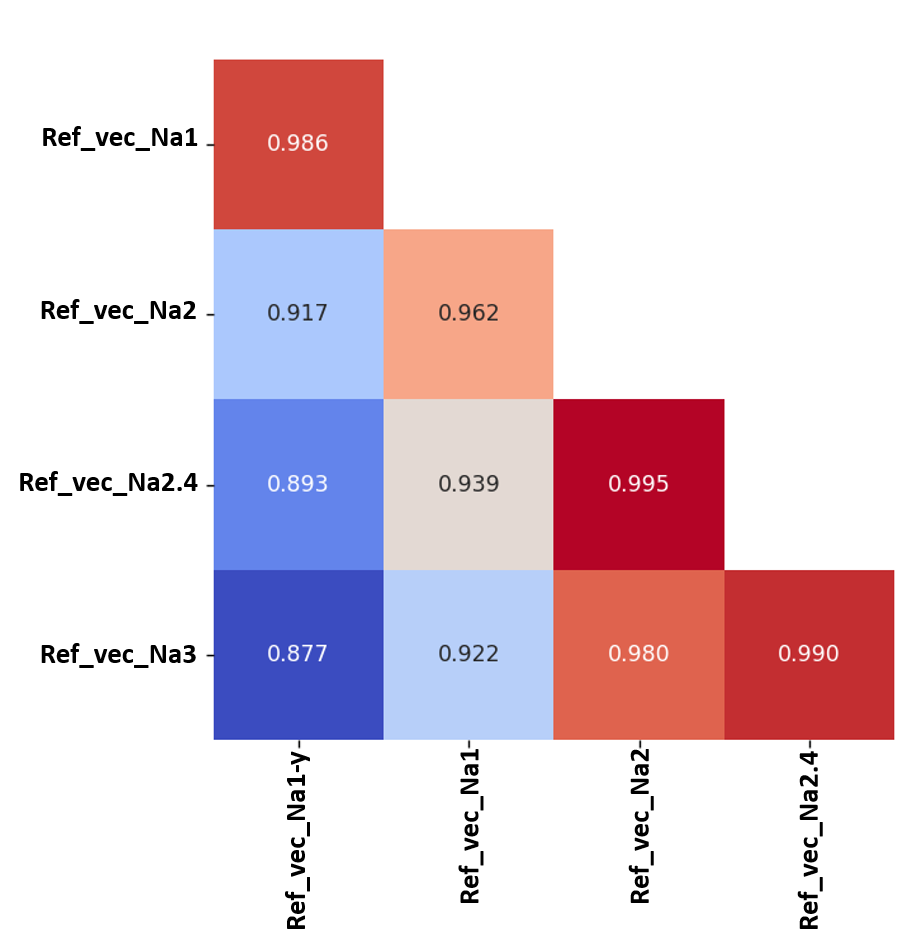}
\caption{Trigonal diagram showing the correlation between the reference vectors using Pearson correlation coefficient.}
\label{fig:supp4}
\end{figure}

\FloatBarrier

\FloatBarrier

\begin{figure}[h!]
\centering
\includegraphics[width=0.95\textwidth]{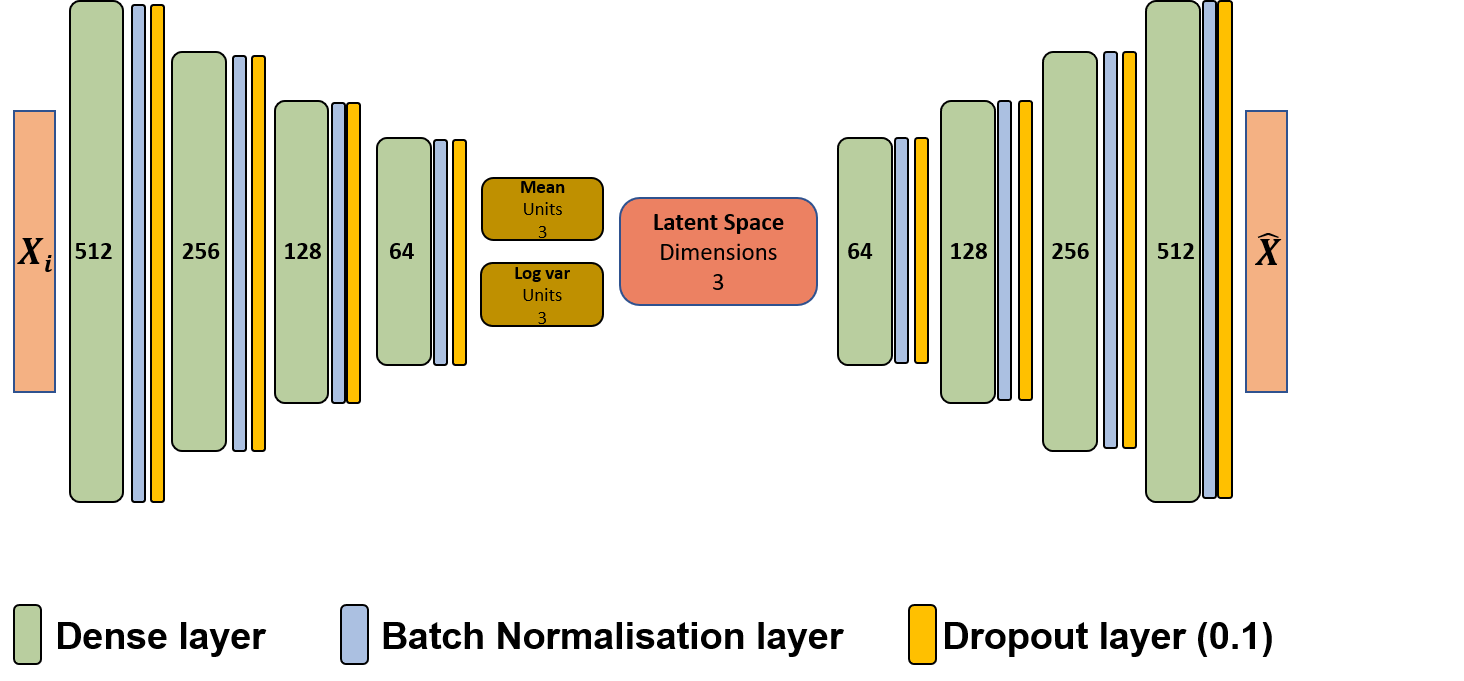}
\caption{Gaussian Mixture Variational AutoEncoder (GMVAE) architecture.}
\label{fig:supp5}
\end{figure}

\FloatBarrier

\begin{figure}[h!]
\centering
\includegraphics[width=0.95\textwidth]{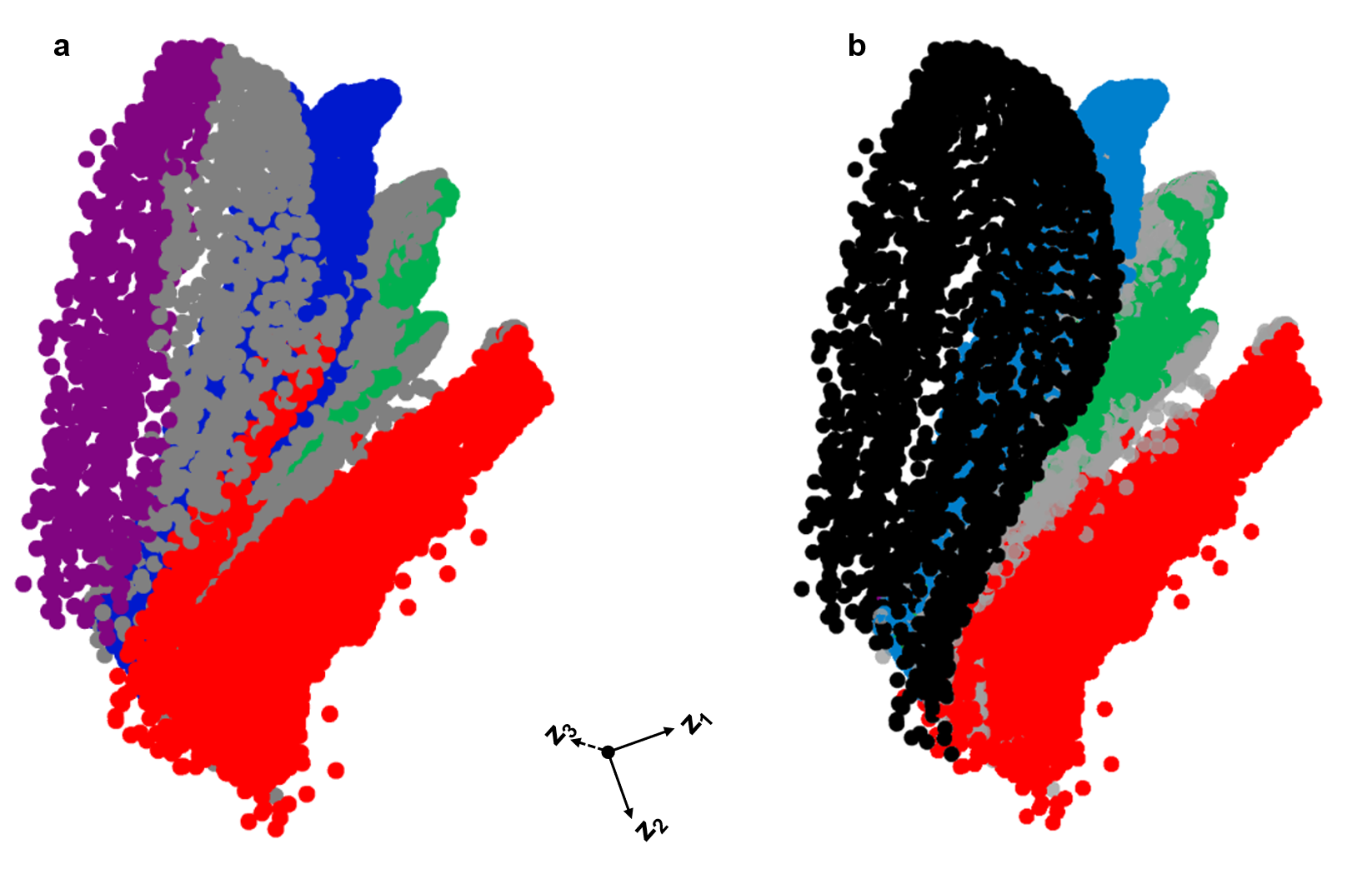}
\caption{Latent space distribution of ten NVPF particles at different states of charge. a) Latent space labeled according to the phase distribution, obtained from Pearson correlation comparison with the reference spectra. b) The projection of the  Na\textsubscript{1}VPF  particle in figure 4.}
\label{fig:supp6}
\end{figure}

\begin{figure}[h!]
\centering
\includegraphics[width=0.95\textwidth]{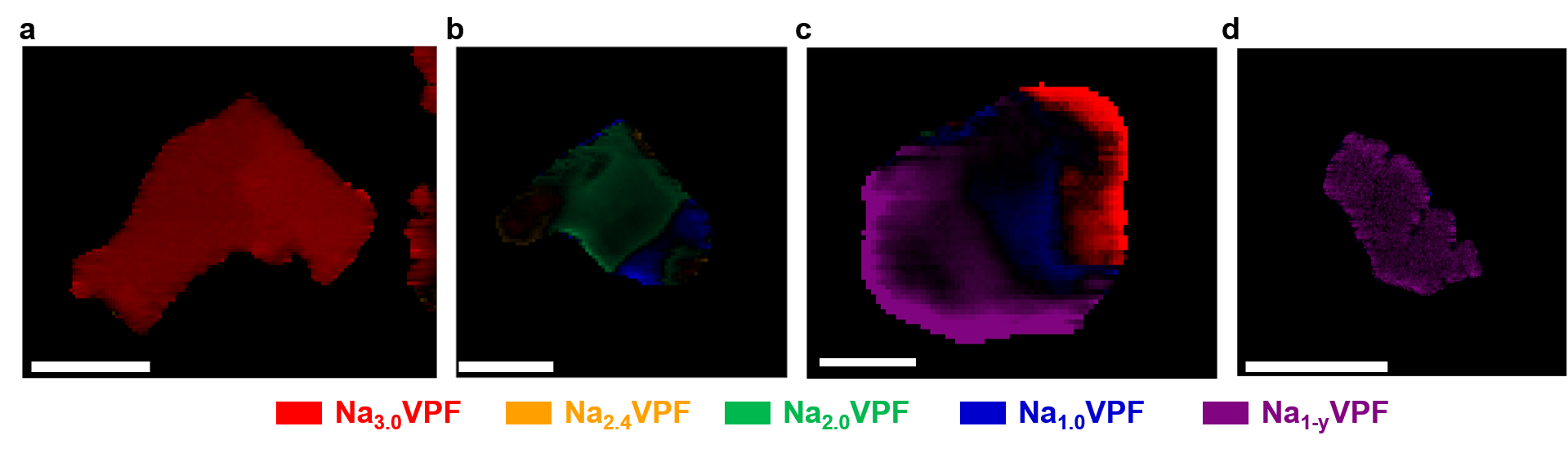}
\caption{(a--d) Reliability maps corresponding to each sample represented in Figure 5, where pixel brightness encodes the confidence in the phase assignment.}
\label{fig:supp7}
\end{figure}

The Pearson correlation quantifies linear similarity between an observed signal and a reference pattern. This approach implicitly assumes that the data can be well described as a scaled and shifted version of the reference. However, in the presence of mixed signals, overlapping phases, or nonlinear spectral distortions, this assumption does not hold. A high correlation value may therefore reflect partial similarity in dominant features or shared variance driven by hidden factors, rather than true phase identity. The projection of the same region into the GMVAE latent space, as shown in Figure \ref{fig:supp6}, revealed that these pixels did not group with the cluster associated with the pristine phase. Unlike correlation analysis performed in the original observation space, the GMVAE learns a structured latent representation that captures multimodality and nonlinear relationships in the data. Phase assignment is then based on probabilistic clustering within this learned manifold, rather than on linear similarity to a single reference pattern. The latent space representation showed that the region occupied an intermediate position between clusters, indicating that the original classification resulted from a linear projection artifact. Consequently, we reclassified this area as an ambiguous region rather than a confidently assigned pristine phase. This ambiguity was subsequently resolved using the GMVAE latent space representation, which provides phase discrimination grounded in the full probabilistic structure of the data and reduces false positive assignments arising from correlation-based analysis. For completeness, the reliability maps derived from the Pearson correlation analysis are included here (Figure \ref{fig:supp7}c the red region) to illustrate the origin of the initial misclassification, while the final resolved phase map obtained from the GMVAE framework is presented in the main Figure 5.

\FloatBarrier

\FloatBarrier

\begin{figure}[h!]
\centering
\includegraphics[width=0.95\textwidth]{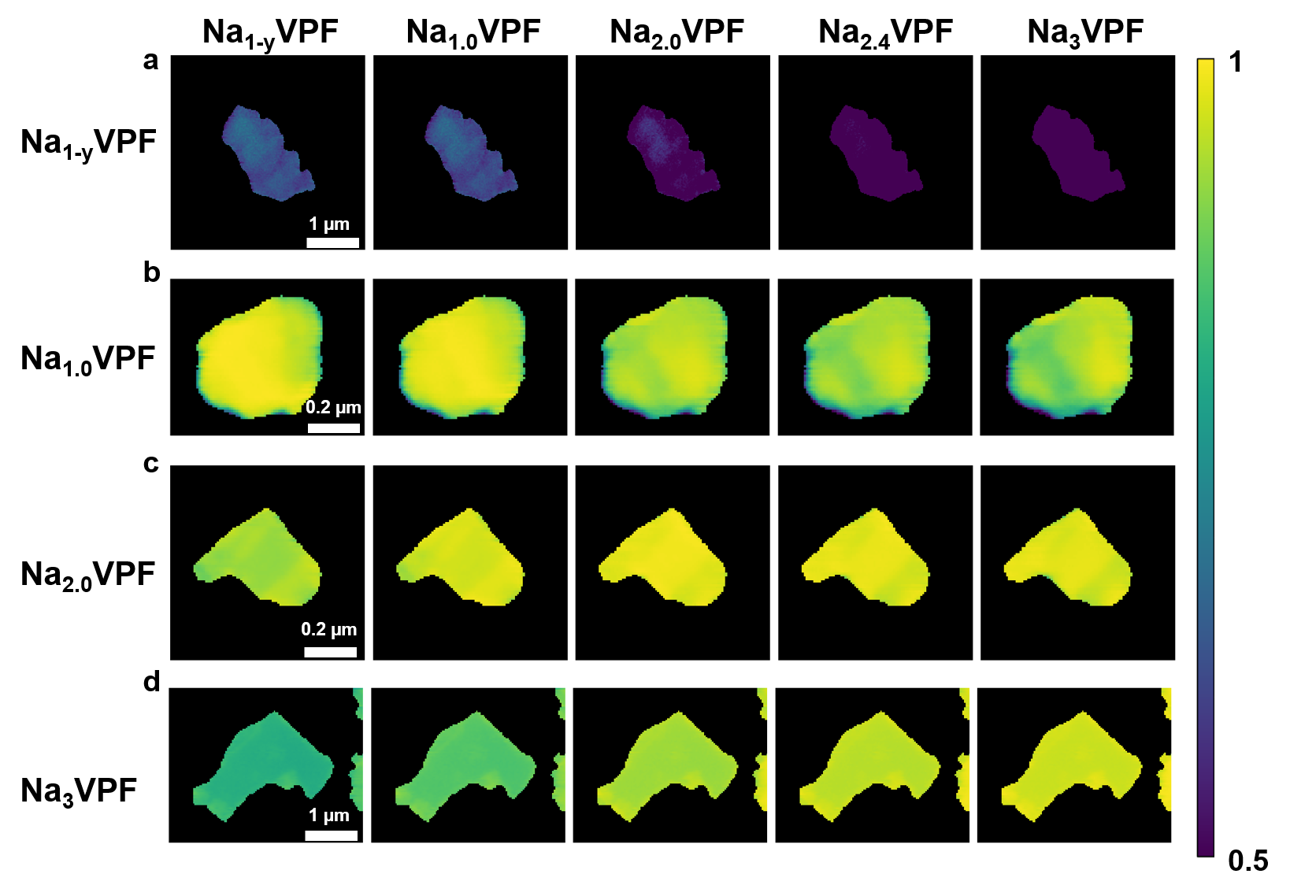}
\caption{Correlation maps of the individual sodium phases associated with the particles presented in Figure 5, from Na\textsubscript{1-y}VPF on the left to Na\textsubscript{3}VPF on the right (as indicated above). The expected composition is shown to the left of the corresponding particle.}
\label{fig:supp8}
\end{figure}
\FloatBarrier

\begin{figure}[h!]
\centering
\includegraphics[width=0.95\textwidth]{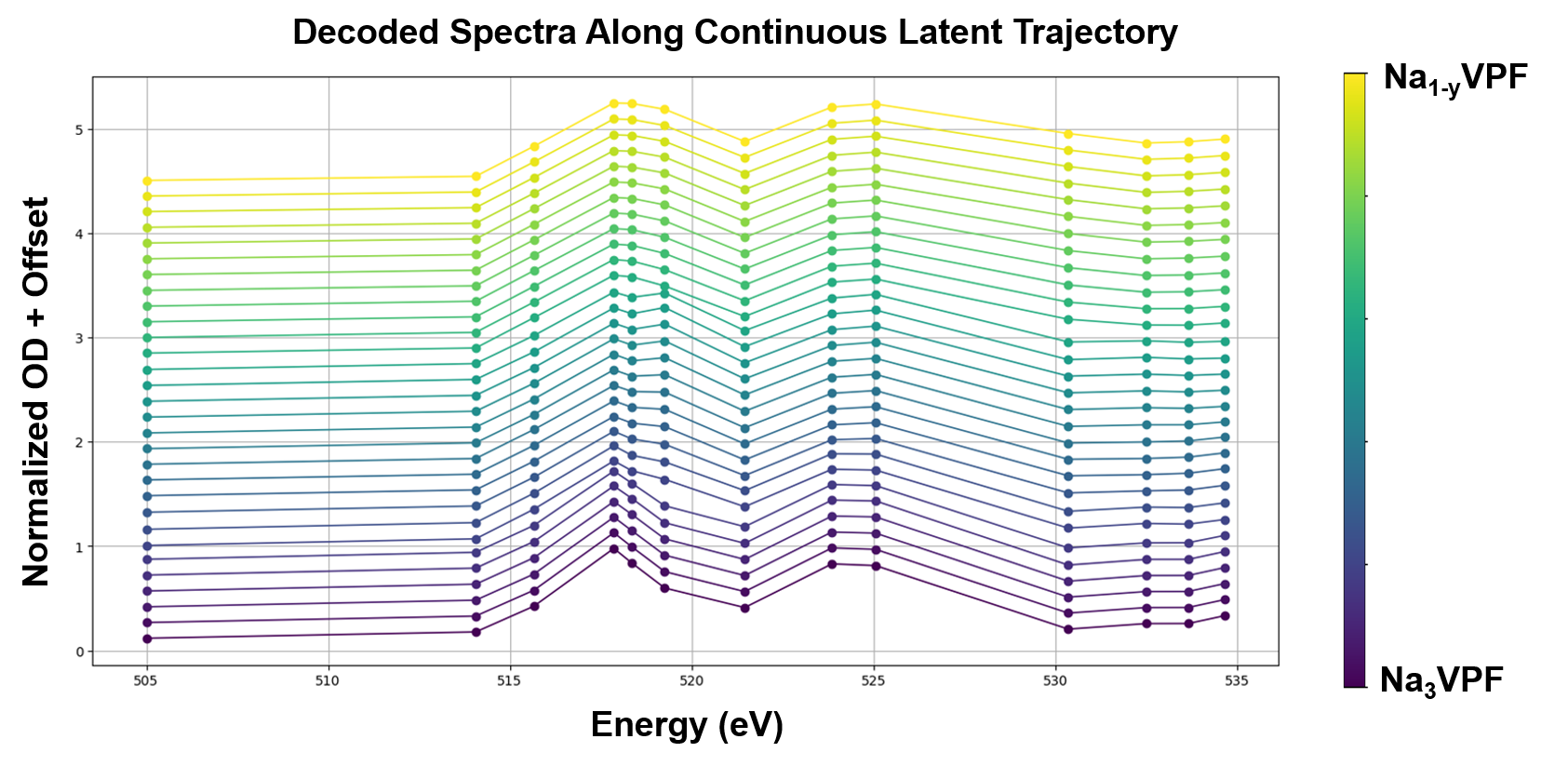}
\caption{The reconstructed X-ray absorption spectra obtained using the decoder only architecture from the projected points in the latent space, spanning the Na\textsubscript{3}VPF phase to the Na\textsubscript{1-y}VPF phase.}
\label{fig:supp9}
\end{figure}

\FloatBarrier

\begin{figure}[h!]
\centering
\includegraphics[width=0.95\textwidth]{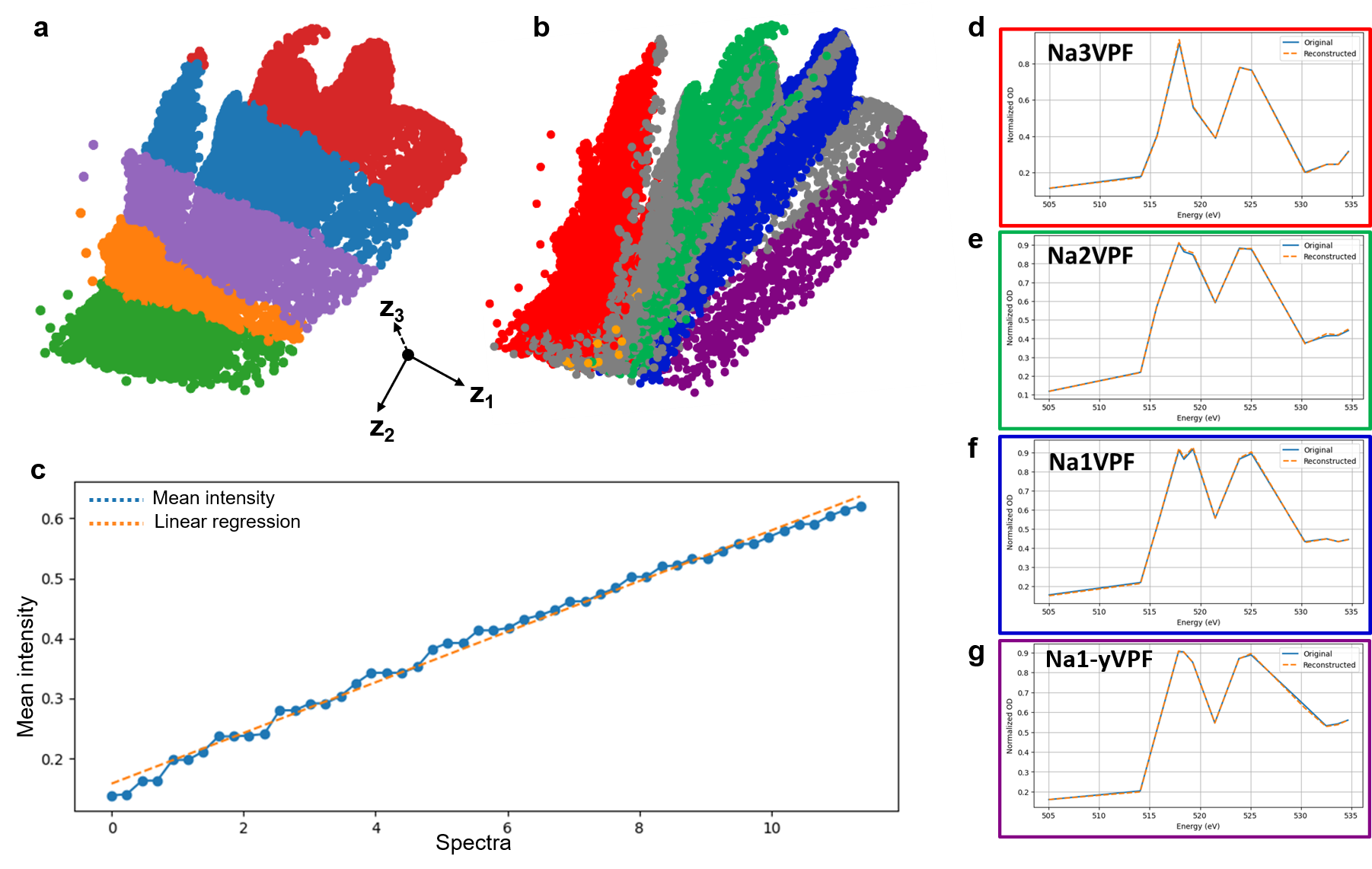}
\caption{Latent space dimension analysis: 
(a) K-means clustering of spectra, with each color representing a different cluster. 
(b) Latent space representation labeled by Pearson correlation, including ambiguous points. 
(c) Mean intensity of the original spectra along a trajectory following the $-z_2$ direction in the latent space. 
Original and reconstructed spectra for each identified phase: 
(d) Na$_{3.0}$VPF region, 
(e) Na$_{2.0}$VPF region, 
(f) Na$_{1.0}$VPF region, 
(g) Na$_{1-y}$VPF region.}
\label{fig:supp10}
\end{figure}
The K-means clustering applied in the latent space reveals distinct clusters, which primarily reflect variations in spectral intensity. In Figure \ref{fig:supp10}a, the identification is based on intensity, with each color corresponding to nearly the same intensity. By calculating the average intensity of each spectrum (each point represents a spectrum) from bottom to top along the $-z_2$ direction, we observe an almost linear increase (Figure \ref{fig:supp10}c). According to the Beer-Lambert law, this trend is directly related to thickness, the thicker the material, the more it absorbs, and the higher the spectral intensity. The different phases are encoded along the remaining latent space axes, as evidenced by the reconstructed spectra corresponding to each cluster (Figure \ref{fig:supp10}d--g). Each reconstructed spectrum closely matches the original (experimental spectrum), demonstrating that the latent space effectively separates the phases while simultaneously capturing intensity (thickness) variations along $z_2$. This visualization highlights the dual functionality of the latent space, one axis ($z_2$) tracks intensity or thickness variations, whereas the other axes encode phase specific spectral features.
\FloatBarrier

\begin{figure}[h!]
\centering
\includegraphics[width=0.95\textwidth]{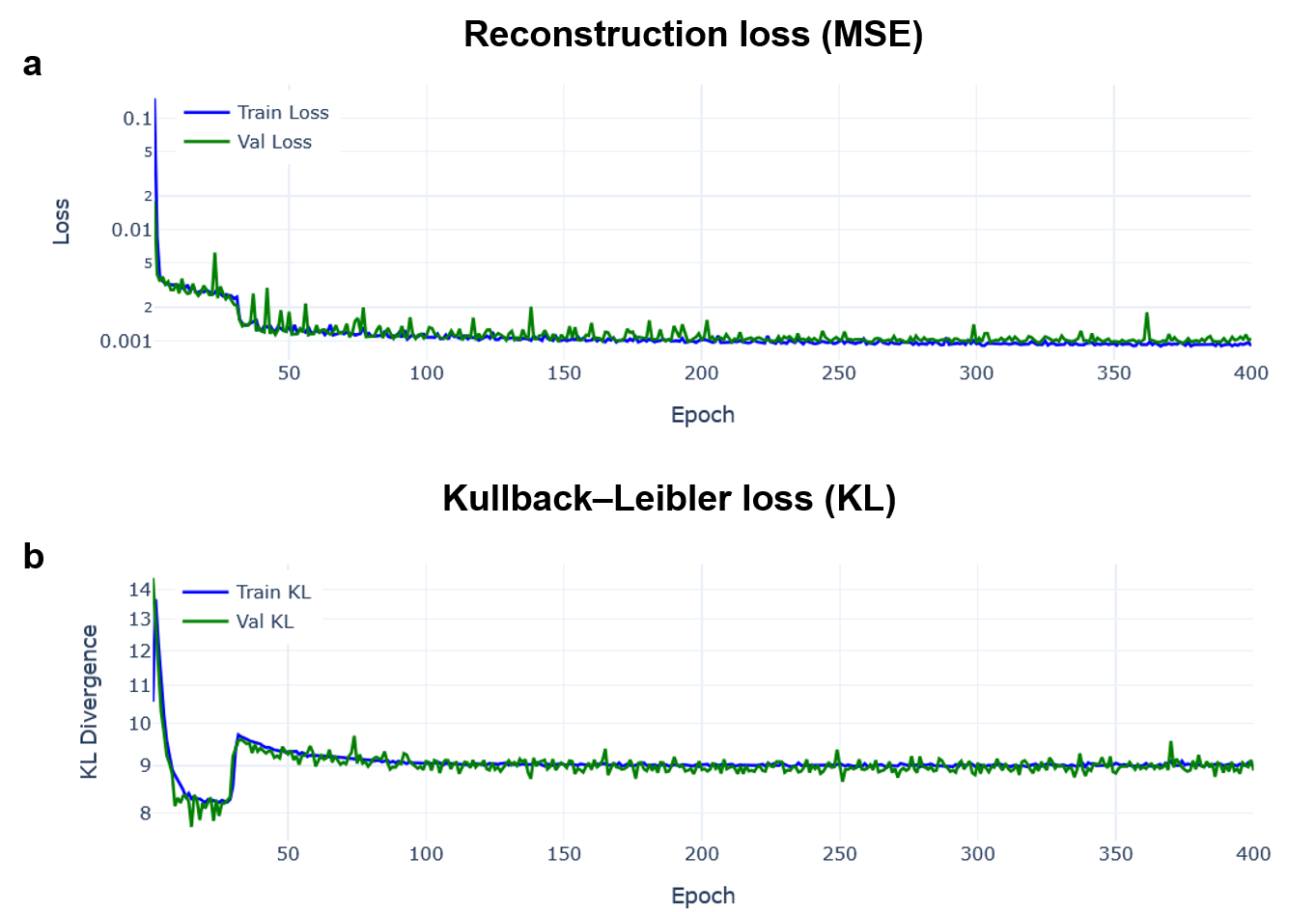}
\caption{Evolution of the mean squared error and Kullback–Leibler losses for the VAE across the training and validation as a function of the epoch, shown on a logarithmic scale.}
\label{fig:supp11}
\end{figure}

\FloatBarrier

\FloatBarrier

\newpage

\section{Algorithms}
\addcontentsline{toc}{section}{Algorithms}

\begin{algorithm}[H]
\caption{Phase Mapping using Pearson Correlation with Ambiguity Tracking}
\label{alg:phase_mapping}
\KwIn{Reference set $R = \{r_m\}_{m=1}^M$, image set $I = \{I_n\}_{n=1}^N$, correlation threshold $\delta = 0.005$}
\KwOut{Phase map $P$, reliability map $L$, ambiguity mask $A$, ambiguous pairs $P_\text{amb}$}

\For{each image $I_n \in I$}{
    Convert $I_n$ to vector $v_n \in \mathbb{R}^p$\;
}
Construct $X \in \mathbb{R}^{p \times N}$ such that $X[i,n] = v_n[i]$\;
\For{each reference $r_m \in R$}{
    Compute correlations $C_m[i]$ using Pearson coefficient:\;
    \[
    C_m[i] = \frac{\sum_{n=1}^{N} (r_m[n]-\bar r_m)(X[i,n]-\bar X_i)}
    {\sqrt{\sum_{n=1}^{N} (r_m[n]-\bar r_m)^2} \sqrt{\sum_{n=1}^{N} (X[i,n]-\bar X_i)^2}}
    \]
}
\For{each pixel $i$}{
    Let $c_i = [C_1[i],\dots,C_M[i]]$\;
    Find best and second-best correlations and their indices\;
    Compute reliability $L[i] = \text{best} - \text{second-best}$\;
    Mark ambiguity $A[i] = 1$ if $L[i] < \delta$\;
    Store ambiguous pair $(i,\text{best phase},\text{second phase})$\;
}
\Return{$P,L,A,P_\text{amb}$}
\end{algorithm}

\begin{algorithm}[H]
\caption{Training GMVAE on XANES vectors using ADAM}
\label{alg:gmvae_lit}
\KwIn{Data $X = \{x_i\}_{i=1}^N \subset \mathbb{R}^d$, batch size $B=128$, learning rate $\eta=10^{-3}$, KL weight $\beta=10^{-5}$, dropout $p_\text{drop}=0.1$, Gaussian components $K=5$, number of training epochs $T=400$}
\KwOut{Trained GMVAE parameters $(\phi, \theta)$}

\For{epoch = 1 \KwTo $T$}{
    Sample a batch $X_B \subset X$ of size $B$\;
    \For{each $x_i \in X_B$}{
        Encode: $z_\text{mean}, z_\text{logvar} = f_\phi(x_i)$ \;
        Reparameterize: $z_i = z_\text{mean} + \exp(0.5\, z_\text{logvar}) \odot \epsilon$, $\epsilon \sim \mathcal{N}(0,I)$\;
        Decode: $\hat{x}_i = g_\theta(z_i)$\;
    }
    Compute batch loss:
    \[
    \mathcal{L}_B = \frac{1}{B}\sum_{i\in X_B} \underbrace{\|x_i - \hat{x}_i\|_2^2}_{\text{reconstruction}} 
    + \beta \underbrace{D_\mathrm{KL}\Big(q_\phi(z_i|x_i) \,\|\, \sum_{k=1}^K \pi_k \mathcal{N}(z_i|\mu_k,\Sigma_k)\Big)}_{\text{GM prior KL}}
    \]
    Update $(\phi, \theta)$ using ADAM optimizer with learning rate $\eta$\;
}
\Return{Trained GMVAE $(\phi, \theta)$}
\end{algorithm}

\printbibliography
